\documentclass[10pt,journal,compsoc]{IEEEtran}
\usepackage[nocompress]{cite}
\usepackage{comment}
\usepackage{graphicx}
\usepackage{fancyvrb}
\usepackage{flushend}
\usepackage[T1]{fontenc}
\usepackage[fleqn]{amsmath}
\usepackage{amsthm}
\usepackage{amssymb,amsfonts}
\usepackage{url}
\usepackage{textcomp}
\usepackage{xcolor}
\usepackage{enumitem}
\usepackage{booktabs}
\usepackage{relsize}
\usepackage{fp}
\usepackage{pifont}
\usepackage{float}

\theoremstyle{definition}
\newtheorem{definition}{Definition}[section]

\RecustomVerbatimEnvironment{verbatim}{Verbatim}{fontsize=\small,fontfamily=zi4,xleftmargin=3mm}

\DeclareMathOperator{\argmax}{\mathrm{arg\,max}}

\DeclareMathOperator{\choices}{\mathit{choices}}
\DeclareMathOperator{\best}{\mathit{best}}

\newcommand{\ident}[1]{\textnormal{\sffamily #1}}
\newcommand{\Din}{X}
\newcommand{\Dout}{Y}

\def\T{T}

\newcommand{\parag}[1]{\vspace{0.75em}\noindent\emph{\sffamily #1.}~}
\newcommand{\email}[1]{#1}

\begin{document}
\title{
\makebox[0pt][l]{\raisebox{0.57in}[0pt][0pt]{\hspace{-1.6in}
\parbox{7.2in}{\footnotesize Final manuscript version of paper accepted for
   publication in  \emph{IEEE Transactions on Knowledge and
   Data Engineering.}\\ \\[-0.75em]
   \copyright 2022 IEEE.  Personal use of this material is permitted.
   Permission from IEEE must be obtained for all other uses, in any current or future media, including reprinting/republishing this material for advertising or promotional purposes, creating new collective works, for resale or redistribution to servers or lists, or reuse of any copyrighted component of this work in other works.
   }}}
AI Assistants: A Framework for Semi-Automated Data Wrangling}

\author{Tomas~Petricek, Gerrit~J.J.~van~den~Burg, Alfredo~Naz{\'{a}}bal, Taha~Ceritli,\\
  Ernesto~Jim\'enez-Ruiz, Christopher~K.~I.~Williams
\IEEEcompsocitemizethanks{%
\IEEEcompsocthanksitem T. Petricek (\email{tomas@tomasp.net}), Charles University, Prague, Czechia\\ (work done while at University of Kent and The Alan Turing Institute)
\IEEEcompsocthanksitem G.J.J. van den Burg (\email{gertjanvandenburg@gmail.com}), Amazon, UK\\ (work done prior to joining Amazon, in The Alan Turing Institute)
\IEEEcompsocthanksitem A. Nazabal (\email{alfredonazabal@gmail.com})\\Amazon Development Centre Scotland, Edinburgh\\(work done prior to joining Amazon, in The Alan Turing Institute).
\IEEEcompsocthanksitem T. Ceritli (\email{taha.ceritli@eng.ox.ac.uk}), University of Oxford, UK\\ (work done in University of Edinburgh and The Alan Turing Institute)
\IEEEcompsocthanksitem E. Jim\'enez-Ruiz (\email{ernesto.jimenez-ruiz@city.ac.uk})\\City, University of London, UK and University of Oslo, Norway
\IEEEcompsocthanksitem C. K. I. Williams (\email{ckiw@inf.ed.ac.uk})\\University of Edinburgh and The Alan Turing Institute, UK} }

\markboth{}{Petricek \MakeLowercase{\textit{et al.}}: AI Assistants}

\IEEEtitleabstractindextext{
\begin{abstract}
Data wrangling tasks such as obtaining and linking data from various sources, transforming data
formats, and correcting erroneous records, can constitute up to 80\% of typical data engineering work.
Despite the rise of machine learning and artificial intelligence, data wrangling remains a tedious
and manual task. We introduce \emph{AI assistants}, a class of semi-automatic interactive tools to
streamline data wrangling. An AI assistant guides the analyst through a specific data wrangling
task by recommending a suitable data transformation that respects the constraints
obtained through interaction with the analyst.

\qquad We formally define the structure of AI assistants and describe how existing tools that
treat data cleaning as an optimization problem fit the definition.
We implement AI assistants for four common data wrangling tasks and make AI assistants easily
accessible to data analysts in an open-source notebook environment for data science, by leveraging
the common structure they follow. We evaluate our AI assistants both quantitatively and
qualitatively through three example scenarios. We show that the unified and interactive design makes
it easy to perform tasks that would be difficult to do manually or with a fully automatic tool.
\end{abstract}

}

\maketitle
\IEEEdisplaynontitleabstractindextext
\IEEEpeerreviewmaketitle

\IEEEraisesectionheading{\section{Introduction}\label{sec:introduction}}
\IEEEPARstart{W}{hile} most \emph{research} in data science focuses on novel methods
and clever algorithms, the \emph{practice} is dominated by the realities of working
with messy data. Surveys \cite{crowdflower2016data,kaggle2017state} indicate
that up to 80\% of data engineering is spent on \emph{data wrangling}, a tedious process of
transforming data into a format suitable for analysis, which includes parsing, making sense
of encodings, merging datasets, and correcting erroneous records. Data wrangling prevents
both organizations and individuals from applying machine learning and represents an
enormous cost, both in terms of wasted time and in terms of missed opportunities.

Despite attempts to address this issue \cite{he2021automl,autods}, data wrangling remains hard to automate,
because it often involves special cases that require human insight. An automatic tool can easily
confuse interesting outliers for uninteresting noise in cases where a human would immediately spot
the difference. This makes incorporating human understanding into the process crucial.
A major advance in the practice of data wrangling therefore requires semi-automated tools that integrate
automatic methods with human insight, allow the analyst to review cleaning operations before applying
them, and follow a unified interface that makes it easy to use a wide range of tools during
data wrangling.

\subsection{Background}
Data wrangling is most often done manually using a combination of programmatic and graphical tools.
Jupyter and RStudio are popular environments used for programmatic data cleaning.
They are used alongside libraries that implement specific functionality such as parsing CSV files
or merging datasets \cite{ccsv,datadiff} and general data transformation
functions provided, e.g.,~by Pandas \cite{pandas} and Tidyverse~\cite{tidyverse}.

Trifacta \cite{trifacta} and OpenRefine \cite{openrefine} are complete graphical data wrangling systems that
consist of myriad tools for importing and transforming data, which are accessible
through different user interfaces or through a scriptable programmatic interface.
Finally, spreadsheet applications such as Excel and business intelligence tools like Tableau \cite{tableau} are
often used for manual data editing, reshaping, and especially visualization \cite{kandel2011research}. The
above general-purpose systems are frequently complemented by ad-hoc tools such as Tabula \cite{tabula},
which extracts tables from PDF documents.

\subsubsection{Semi-automatic data wrangling}
Some of the most practical tools along the entire data wrangling pipeline partially automate
a specific tedious data wrangling task. To merge datasets, Trifacta \cite{trifacta} and datadiff
\cite{datadiff} find corresponding columns using machine learning. To transform textual
data and tables, Excel \cite{excelbyexample} employs programming-by-example to parse semistructured
data, LearnPADS \cite{learnpads} automatically generates programmatic data processing routines,
and many tools exist to semi-automatically detect duplicate records in databases~\cite{duplicate}.

A common theme in data wrangling tools that utilize machine learning, including those listed
above, is that they allow the analyst to review and influence the results. The interaction
between a human and a computer in such data wrangling systems follows a number of common patterns:

\begin{itemize}
\setlength\itemsep{0.25em}
\item \emph{Onetime interaction.} A tool makes a best guess, but allows the analyst to manually edit
the proposed data transformation. Examples include LearnPADS \cite{learnpads} and dataset merging
in Trifacta \cite{trifacta} and datadiff \cite{datadiff}.

\item \emph{Live previews.} Environments like Jupyter, Trifacta \cite{trifacta}, and The Gamma
\cite{livegamma} provide live previews, allowing the analyst to check the results and tweak
parameters of the operation they are performing before moving~on.

\item \emph{Iterative.} A tool re-runs inference after each interaction with a human
to refine the result. For example, in Predictive Interaction \cite{heer2015predictive}
the analyst repeatedly selects examples to construct a data~transformation.

\item \emph{Question-based.} A system repeatedly asks the human questions about data and uses the
answers to infer and refine a general data model. Examples include data repair tools such as UGuide
\cite{repair,uguide}.
\end{itemize}

\noindent
The interaction pattern that combines human inputs and automatic inference is also known as
mixed-initiative interfaces \cite{mixedinitiative,williams2020units} in the context of
graphical user interfaces, and as human-in-the-loop data analytics (HILDA)
\cite{doan2018human,williams2020units} in the context of data science.
However, both of these are general patterns, rather than  specific technical frameworks.

\subsubsection{Issues and limitations}
The emerging class of semi-automatic data wrangling tools have the potential to dramatically
simplify data wrangling because they combine the automation and scalability of machine
learning with crucial human insight. However, this development has been hindered by two main issues.

First, semi-automatic data wrangling tools lack a common structure. Notions such as
mixed-initiative user interfaces and human-in-the-loop are too general and do not provide a
specific technological framework that a tool implementation could follow. Moreover,
many tools only exist in one specific environment or programming language, forcing the
analyst to repeatedly switch between tools. They may, for example, need to export data from
Trifacta to a CSV file, run a particular R or Python script and then import data back. This
is not without risk, as intermediate data formats may accidentally corrupt data.

Second, the way analysts interact with such tools can vary significantly. Consequently, users
have to learn how to interact with each new tool using whatever mechanism it supports, be it a
graphical user interface, a program library, or a command-line script. Moreover,
most semi-automatic data wrangling tools accept only limited forms of human input.
The \emph{onetime interaction} pattern of interaction prevails and only a few systems
\cite{heer2015predictive,guo2011proactive} follow the flexible \emph{iterative} pattern.
Even then, the way of specifying feedback in such systems is often specialized and tied to
the problem domain.

\subsection{Contributions}
We present the notion of an \emph{AI assistant}, a common structure for building
semi-automatic data wrangling tools that incorporate human feedback. AI assistants capture
the \emph{iterative} pattern of interaction where a human user repeatedly provides insights about
the problem and a computer performs automatic inference. The design addresses the issues with
semi-automatic data wrangling tools described above.

First, the AI assistant framework allows for a wide range of semi-automatic data wrangling tools
that can integrate human feedback. For analysts, this makes using AI assistants
easy as they can complete a variety of data wrangling tasks through a uniform user interface.
For tool developers, this makes building AI assistants easier, because any AI assistant can be
readily used from JupyterLab and potentially other data wrangling systems.

Second, the notion of an AI assistant defines a simple uniform mechanism for iteratively providing
feedback to the assistants. An AI assistant makes an initial best guess and then it repeatedly
offers the analyst a list of options that they can choose from in order to guide the next
iteration of the automatic process.

The remainder of this paper is structured as follows:

\begin{itemize}
\setlength\itemsep{0.25em}
\item We introduce AI assistants by example in Section~\ref{sec:motivation}, looking at how
  the datadiff AI assistant simplifies merging data from inconsistent datasets.

\item We define the structure of AI assistants formally in Section~\ref{sec:theory} and show how
  tools solving an optimization problem fit the definition.

\item We present four AI assistants in Section~\ref{sec:aia} (for parsing, merging, type inference, and semantic type prediction),
	that each restructure an existing non-interactive tool as an interactive AI assistant.

\item We evaluate our approach in Section~\ref{sec:eval} qualitatively, by discussing three scenarios
  where automatic tools would fail, and quantitatively, by evaluating how many interactions
  are needed to complete a wrangling~task.
\end{itemize}

\noindent
While we may not entirely eliminate the 80\% of time data scientists spend on data wrangling,
our framework provides a pathway to the future where data analysts leverage the advances in
AI for the most time-consuming aspect of their job. The four AI assistants we develop illustrate
the benefits that a rich ecosystem of AI assistants would provide.


\section{Motivation}
\label{sec:motivation}

To give an overview of how AI assistants work, we discuss the data wrangling task of merging
multiple incompatible datasets, using the UK broadband quality data \cite{ofcom}, published
by the UK communications regulator Ofcom.

The regulator collects data annually, but the formats of the files are inconsistent over the
years. The order of columns changes, some columns are renamed, and new columns are added. We take
the 2014 dataset and select six interesting columns (latency, download and
upload speed, time needed to load a sample page, country, and whether the observation is from
an urban or a rural area). We then want to find corresponding columns in the 2015 dataset.

The 2015 dataset has 66 different columns so finding corresponding columns manually would be
tedious. Instead, we can use the automatic datadiff tool~\cite{datadiff}, which matches columns
by analyzing the distributions of the data in each column. Datadiff generates a list of
\emph{patches} that reconcile the structure of the two datasets. A patch describes a single data
transformation to, for example, reorder columns or recode a categorical column according to an
inferred mapping. Datadiff is available as an R function that takes two datasets and several
hyperparameters that affect the likelihood of the different types of patches.
Datadiff correctly matches five out of six columns, but it
incorrectly attempts to match a column representing Local-loop unbundling (LLU) to a column
representing UK countries. This happens as datadiff allows the recoding of categorical columns,
and seeks to match them based on the relative frequencies in the two columns. Consequently,
the inferred transformation includes a patch to recode the \texttt{Cable}, \texttt{LLU}, and
\texttt{Non-LLU} values to \texttt{Scotland}, \texttt{Wales}, and \texttt{England}. To correct this,
we could
either manually edit the resulting list of patches, or tweak the likelihood of the
\emph{recode} patch. Such parameter tuning is typical for real-world data wrangling, but finding
the values that give the desired result can be hard.

\begin{figure}
\includegraphics[scale=0.158]{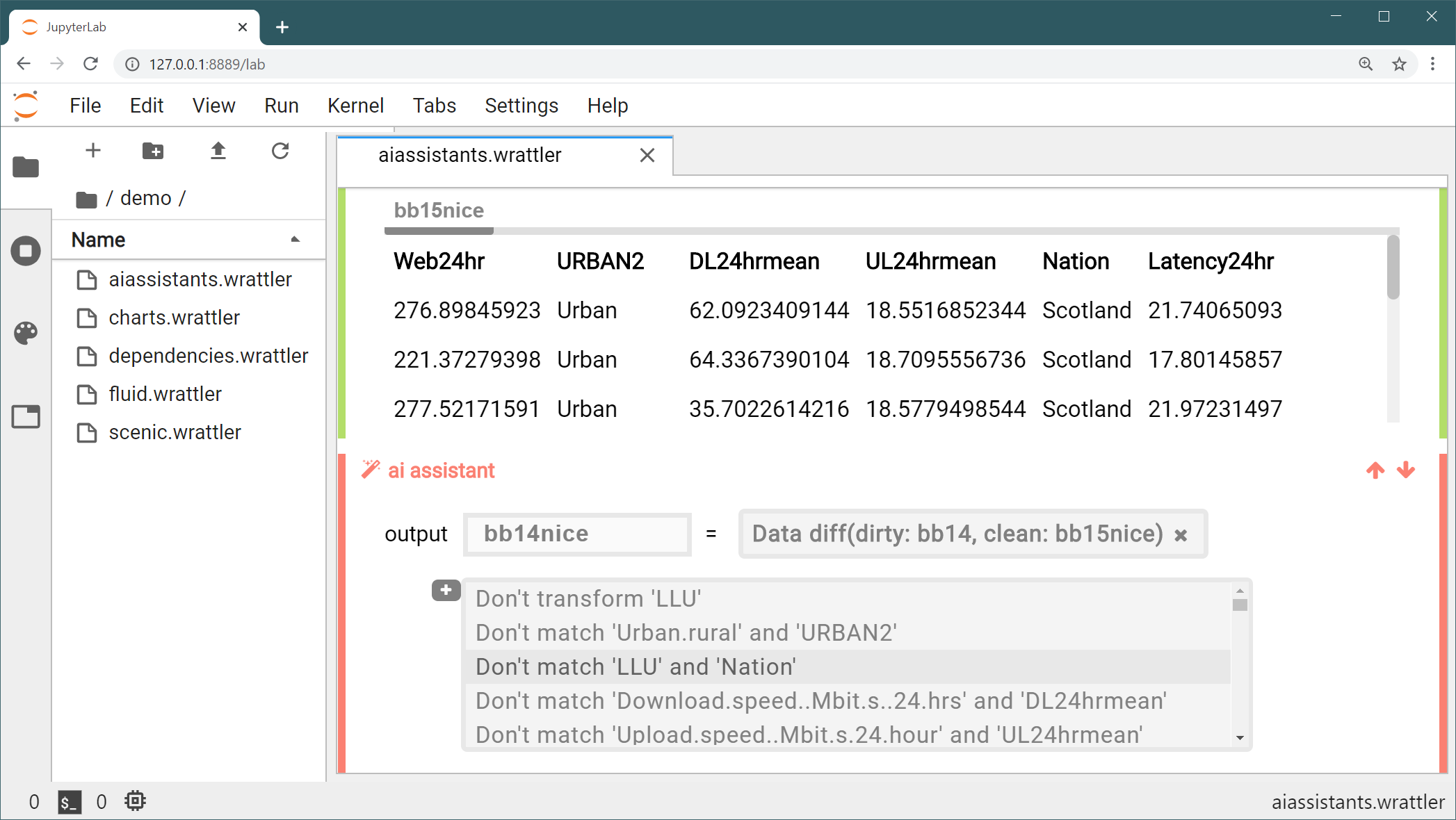}
\caption{Using the datadiff AI assistant in JupyterLab to semi-automa\-ti\-cally merge data
from two sources, parsed by an earlier R script.}
\label{fig:ddiff-jupyter}
\vspace{-0.5em}
\end{figure}

The semi-automatic datadiff AI assistant presented in this paper enables the analyst to guide
the inference process by specifying human insights in the form of constraints.
The AI assistant first suggests an initial set of patches with one incorrect mapping. After the
analyst chooses one of the offered constraints, shown in Figure~\ref{fig:ddiff-jupyter}, datadiff
runs again and presents a new solution that respects the specified constraints
until, after two more simple interactions, it reaches the correct solution (see
Section~\ref{sec:eval-ddiff} for details).

The example illustrates the interaction pattern at the core of AI assistants. At each step,
the assistant analyzes the data and recommends the best data transformation. It previews the transformed
data and offers a range of constraints that may be sorted by their estimated fit. The analyst
can then accept the result or choose another constraint to refine the outcome. The
behaviour is controlled through comprehensible constraints rather than opaque numerical parameters.

As discussed in Appendix~\ref{sec:system} (see supplemental files),
we make AI assistants available in JupyterLab, which allows analysts to combine text
and equations with code and outputs such as charts. We introduce a new cell
type that leverages the common structure of AI assistants to provide a unified user interface
(see Figure~\ref{fig:ddiff-jupyter}) for accessing any AI assistant.


\section{Theory of AI assistants}
\label{sec:theory}

The notion of an \emph{AI assistant} formally captures a pattern of interaction between a semi-automatic
data wrangling tool and a data analyst. The precise definition distinguishes AI assistants from more
general notions such as human-in-the-loop data analytics, and it facilitates the development of
concrete AI assistants discussed in Section~\ref{sec:aia}.

\subsection{Formal model}
\label{sec:theory-pl}

Our definition uses the algebraic approach \cite{bird96algebra}
and presents AI assistants as a formal mathematical entity that consists of several
operations, modeled as mathematical functions between different sets. For reference,
a glossary of symbols used in the paper can be found in Table~\ref{tab:symbols} (supplement).

Every AI assistant is defined by three operations that work with expressions~$e$, past human
interactions $H$, input data $\Din$, and output data $\Dout.$ While AI assistants share a common
structure, the language of
expressions $e$ that an assistant produces, the notion of human interactions $H$, and the notion of
$\Din$ and $\Dout$ can differ between assistants.

We refer to $e$ as expressions,
following the programming research tradition, but expressions $e$ can also be thought of as data
cleaning scripts. As we will see from our concrete examples, the input data $\Din$ is typically one
or more data tables and the output data $\Dout$ is typically a single table, often annotated with
meta-data such as column types.

\begin{definition}[AI assistant]
\label{def:aia}
Given expressions $e$, input data $\Din$, output data $\Dout$, and human interactions $H$, an
\emph{AI assistant} $(H_0, f, \best, \choices)$ is a tuple where
$H_0$ is a set denoting an empty human interaction and
$f, \best$ and $\choices$ are operations such that:
\begin{itemize}
\item $f(e,\Din) = \Dout$
\item $\best_{\Din}(H) = e$
\item $\choices_{\Din}(H) = (H_1,H_2,H_3,\ldots,H_n)$.
\end{itemize}
\end{definition}

\noindent
The operation $f$ transforms an input dataset $\Din$ into an output dataset $\Dout$ according to the
expression $e$. The operation $\best_{\Din}$ recommends the best expression
for a given input dataset $\Din$, respecting past human interactions $H$. Finally, the operation
$\choices_{\Din}$ generates a sequence of options $H_1, H_2, H_3, \ldots, H_n$ that the
analyst can choose from (for instance through the user interface illustrated in Figure~\ref{fig:ddiff-jupyter}).
When interacting with an assistant, the selected human interaction $H$ is passed back to
$\best_\Din$ in order to refine the recommended expression. Note that the sequence of human
interactions given by $\choices_{\Din}$ may be sorted, starting with the one deemed the most
likely. To initialize this process, the AI assistant defines an empty human interaction $H_0$.

The interesting AI logic can be implemented in either the $\best_\Din$
operation, the $\choices_\Din$ operation, or both. The $f$ operation is typically
straightforward. It merely executes the inferred cleaning script. Both $\best_{\Din}$ and
$\choices_{\Din}$ are parameterized by input data $\Din$, which could be the actual input or a
smaller representative subset, such as coresets~\cite{coresets}, to make working with the
assistant more efficient.

\begin{figure}
\centering
\noindent
\includegraphics[width=0.48\textwidth]{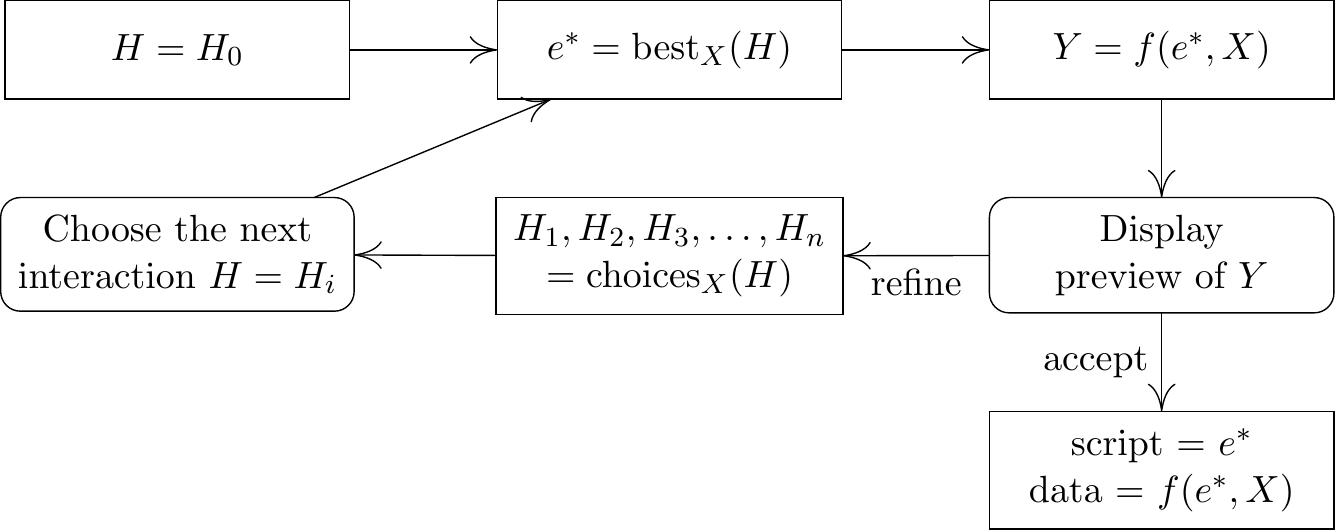}
\begin{minipage}{0.32\textwidth}
  \vspace{-5em}
  \caption{Flowchart illustrating the interaction between an analyst and an AI assistant.
  Steps drawn as rounded rectangles correspond to user interactions with the system.}
  \label{fig:interaction}
\end{minipage}
\begin{minipage}{0.16\textwidth}
~
\end{minipage}
\vspace{-2.5em}
\end{figure}

The logic of working with AI assistants is illustrated in Figure~\ref{fig:interaction}. When using
the assistant, we start with the empty interaction $H_0$. We then iterate until the human analyst
accepts a proposed data transformation. In each iteration, we first invoke $\best_\Din(H)$ to get
the best expression $e^{*}$ respecting the current human insights captured by $H$. We then invoke
$f(e^{*}, \Din)$ to transform the input data $\Din$ according to $e^{*}$ and obtain a transformed
output dataset $\Dout$. After seeing a preview of $\Dout$, the analyst can either accept or reject
the recommended expression $e^*$. In the latter case, we generate a list of possible human
interactions $H_1, H_2, H_3, \ldots, H_n$ using $\choices_\Din(H)$ and ask the analyst to pick an
option $H_i$ (where $i\in \{1,\ldots,n\}$). We use this choice as a new human interaction $H$ and call
the AI assistant again.

Having a unified structure for AI assistants means that we can separate the development of
individual AI assistants from the development of tools that use them. Our JupyterLab implementation
facilitates access to any AI assistant that adheres to the interface captured by
Definition~\ref{def:aia}.

\subsection{Example}
To provide intuition behind the operations, we return to the semi-automatic datadiff AI
assistant introduced in Section~\ref{sec:motivation} and presented in full in Section~\ref{sec:aia-ddiff}.
In case of datadiff, an expression $e$ is a list of patches. Input data $\Din$ is a pair of data
tables comprising a reference dataset and an input dataset. The
output data $\Dout$ is the input dataset, transformed to the format of the reference dataset.
Finally, human interactions $H$ are lists of constraints that restrict what expressions are
permissible. An example constraint, discussed earlier, prevents the matching of particular columns.

The most interesting aspect of the assistant is the $\best_\Din$ operation. It takes a sample input
$\Din$ together with a trace of human interactions $H$, which is a list of constraints. It then
finds the best way to match the columns from the two datasets, utilizing the algorithm
from the original datadiff~\cite{datadiff}, but respecting the constraints. The result is a
list of patches, which is returned as an expression $e$. The $\choices_\Din$ operation
generates a list of choices $H_1, H_2, H_3, \ldots, H_n$. An individual choice is obtained by taking the
constraints specified earlier and adding one additional constraint that restricts some aspect of
the recommended script, e.g.,~recoding of a column that was recommended in the expression $e$.
Finally, $f$ applies the list of patches to the input data.

In datadiff, the clever algorithmics are done in the $\best_\Din$ operation, while $\choices_\Din$
is simpler. It generates constraints in a simple hard-coded way, although a more elaborate AI assistant
could rank these constraints.


\subsection{Optimization perspective}
\label{sec:theory-ml}

Our definition of an AI assistant is purposefully general, but the way
most AI assistants recommend cleaning scripts is based on the optimization of
an objective function. They attempt to find the best cleaning script for a given problem from a set
of possible cleaning scripts. The best script is determined by an objective
function $Q$ that scores expressions based on how well they clean the specified input data.

As before, we write $e$ for expressions (cleaning scripts) that an AI assistant recommends and $H$
for human interactions. The operation $\best_\Din(H)$ solves an optimization problem based on an
objective function $Q_H$ to identify the best expression $e^*$ from a set $E_H$ of possible
expressions. Note that both $Q_H$ and $E_H$ are parameterized by human interactions, meaning that
interaction with the tool can affect both the optimization objective and the set of permitted expressions.
More formally, given $Q_H$ and $E_H$ where:
\begin{itemize}
\item[--] $Q_H(\Din, e)$ is an objective function that assigns a score to an expression
$e$, applied to input data $\Din$, taking into account human interaction $H$;
\item[--] $E_H$ is a set of permitted expressions with respect to human interaction $H$,
\end{itemize}
we define $\best_\Din$ as solving an optimization problem:

\vspace{0.5em}
$\best_\Din(H) = \argmax_{e \in E_H} Q_H(\Din, e)$
\vspace{0.5em}

\noindent
The objective function $Q$ needs to be defined individually for each AI assistant. It typically
uses a measure of how clean the data is after applying the expression $e$.
For datadiff, $Q$ is computed as a sum of distance measures between the empirical distributions of
the corresponding columns~\cite{datadiff}. The optimization based on $Q$ is also implemented
individually for each AI assistant and is discussed in the next section.

The fact that both $E_H$ and $Q_H$ are parameterized by $H$
makes the definition more flexible. One human interaction can entirely prevent the
assistant from generating certain expressions (by removing them from $E_H$), while another can
make a particular expression less desirable (by decreasing the score assigned to it by $Q_H$).
For example, human interactions in datadiff
restrict the set of allowed expressions $E_H$, but do not affect the
objective function $Q_H$.


\section{Practical AI assistants}
\label{sec:aia}

In this section, we show how to turn four existing non-interactive data wrangling tools
into interactive AI assistants. An example of a newly developed assistant for outlier
detection is discussed in Appendix~\ref{sec:aia-outlier} (supplemental files).


\subsection{datadiff: Merging mismatched data tables}
\label{sec:aia-ddiff}

We start by revisiting the datadiff AI assistant. The original R package
\cite{datadiff} implements a function that returns the inferred best list of patches.
We modify the package to support restricting the optimization using constraints specified by
the analyst, and use it as the basis for an interactive AI assistant.
The following formal model explains how the AI assistant uses the underlying
optimization and generates choices that the analyst can use to control the assistant.

\parag{Formal definition of datadiff}
The input data for the assistant is a pair of data tables $\T_i, \T_r$ representing the
input and reference datasets, respectively. The expression $e$ is a sequence of patches and
human interactions $H$ are lists of constraints that restrict what patches can be generated.
Patches $P$ and constraints $c$ are defined as:

\vspace{0.75em}
$\begin{array}{lcll}
 P &=& \ident{recode}(k, [v_1\mapsto v_1', \ldots]) ~|\, \ident{linear}(k, a, b) ~|\\[0.25em]
   && \ident{delete}(k) ~|\, \ident{insert}(k) ~|\, \ident{permute}(\pi)\\[0.25em]
 c &=& \ident{nomatch}(k,l) ~|\, \ident{notransfrom}(k) ~|\, \ident{match}(k,l)
\end{array}$
\vspace{0.75em}

\noindent
The $\ident{recode}(k, [v_1\mapsto v_1', \ldots])$ patch transforms
categorical values in a column $k$ by replacing old values $v_i$ with new values $v_i'$,
$\ident{linear}(k, a, b)$ transforms values $v$ in a numerical column $k$ using a linear
transformation $v \cdot a + b$, $\ident{insert}(k)$ inserts a new column at an index $k$,
$\ident{delete}(k)$ removes a column at an index $k$, and $\ident{permute}(\pi)$ reorders
columns according to a permutation $\pi$.

Interacting with the assistant results in a list of constraints:
$\ident{notransform}(k)$ prevents recoding or linear transformation in a column $k$, while
$\ident{nomatch}(k, l)$ and $\ident{match}(k, l)$ prevent or enforce matching of columns
$k$ from the input dataset to the column $l$ in the reference dataset.

The operations of datadiff follow the optimization-based framework where $\best_\Din(H)$
finds a list of patches from the set $E_H$ that maximizes the objective function $Q$:

\vspace{0.75em}
$\begin{array}{l}
\best_\Din(H) = \argmax_{e \in E_H} Q(\Din, e)~\textnormal{where}\\[0.25em]
\quad E_H = \{ (P_1, \ldots, P_L) \in E ~|~ \forall i\in 1\ldots L.\,\ident{valid}_H(P_i) \}
\end{array}$
\vspace{0.5em}

\noindent
The objective function, discussed below, is not affected by human interaction, but
human interactions do limit the set of allowed expression $E_H$. This is
captured by the $\ident{valid}_H$ predicate defined in Appendix~\ref{app:details} (see supplemental
files). Briefly, a list of patches $P$ is valid if it does not recode or rescale any column specified
in $\ident{norecode}$ and if the permutation given in $\ident{permute}(\pi)$ is compatible with
the $\ident{match}$ and $\ident{nomatch}$ constraints.

The $\choices_\Din$ operation, also given in Appendix~\ref{app:details},
offers constraints based on the best patch set obtained from calling
$\best_\Din$. Each human interaction adds one additional constraint to the current set of
constraints $H$. The constraints allow the analyst to override some aspect of the generated patch.
For any $\emph{recode}$ or $\emph{linear}$ patch, we offer the $\ident{notransform}$ constraint to
block the transformation. For any matched columns, we offer $\ident{nomatch}$, and for any columns
that were not automatically matched we offer the $\ident{match}$ constraint to manually match them.
In the example discussed above, the assistant recommends an incorrect $\ident{recode}$ patch.
The first interaction offered by $\choices_\Din$ is to add the $\ident{notransform}$ constraint
to prevent this matching.


\parag{Objective function optimization}
We use the same objective function $Q$ as non-interactive datadiff \cite{datadiff}.
Given the input and reference datasets and a set of patches to apply, the objective function
sums the distances between the distributions of the matched columns, using
the Kolmogorov-Smirnov statistic for numerical columns and the total variation (TV) statistic
for categorical columns.

The optimization algorithm employed in datadiff first computes the optimal
patch for all pairs of columns producing a cost matrix. The optimal matching is then determined
by running the Hungarian algorithm \cite{kuhn1955hungarian}. Our modification incorporates
the constraints specified by the user by not applying recoding where prevented by a constraint and
by setting the cost of columns that should or should not be matched to zero and infinity,
respectively. Details and performance considerations can be found in Appendix~\ref{app:performance}.

\parag{Example of using datadiff}
Suppose we have two data tables and we want to transform the input table $T_i$ on the
left to match the format of the reference table $T_r$ on the right. The following shows the header
and the first three rows:

\vspace{0.5em}
\begin{verbatim}
City, Name, Count          Name, City
Cardiff, Alice, 1          Joe, London
Cardiff, Bob, na           Jane, Edinburgh
Edinburgh, Bill, 2         Jim, London
\end{verbatim}
\vspace{0.5em}

\noindent
The original datadiff recommends three patches: $\ident{delete}(3)$, $\ident{permute}(2,1)$
and $\ident{recode}(2, [\textnormal{``Cardiff''}\mapsto\textnormal{``London''}])$.

Datadiff correctly infers that we need to drop the {\ttfamily Count} column and that
the order of {\ttfamily Name} and {\ttfamily City} has been switched. It erroneously infers
that the encoding of a categorical column {\ttfamily City} has been changed.
This would be useful for pairs of values like ``true'', ``false'' and ``yes'', ``no'', but it
is incorrect in the case of cities.

Using the interactive datadiff, the analysts can specify the $\mathit{notransform}(2)$ constraint,
which will prevent datadiff from generating the $\ident{recode}$ patch for the column $2$. The
interactive AI assistant makes such an intervention easy, because it offers the constraint via the
$\choices_\Din$ operation and the analyst can simply select it from a drop-down menu.


\subsection{CleverCSV: Parsing tabular data files}
\label{sub:clevercsv}

While parsing CSV files in the standard format \cite{rfc4180} is easy, parsing a file with
non-standard column separators and other formatting parameters often requires human insight.
CleverCSV~\cite{ccsv} is an automatic tool that uses a data consistency measure to
determine formatting parameters, called a ``dialect'', consisting of the delimiter (e.g.,~\verb+,+),
quote (e.g.,~\verb+"+) and escape characters (e.g.,~\verb+\+). We adapt CleverCSV into an interactive
AI assistant that allows the analyst to guide the tool in case the automatic detection fails.

\parag{Formal definition of CleverCSV}
CleverCSV is an optimi\-zation-based assistant that takes a single string, representing the CSV file,
as the input data $\Din$. The objective function $Q(X, e)$ is defined by the data
consistency measure discussed below, expressions $e$ represent
dialects, and $Y = f(e, \Din)$ denotes the result of parsing the file using a given dialect.
Human interactions place constraints on the characters that are considered for each parameter
and can either fix a dialect parameter to a specific value or block a character from being considered:

\vspace{0.75em}
$\begin{array}{lcll}
	e &=& (\ident{is\_delimiter}(d), \ident{is\_quote}(q), \ident{is\_escape}(a)) \\[0.25em]
	c &=& \ident{fix\_delimiter}(d) ~|~ \ident{fix\_quote}(q) ~|~ \ident{fix\_escape}(a)\\[0.25em]
  &|& \ident{not\_delimiter}(d) ~|~ \ident{not\_quote}(q) ~|~ \ident{not\_escape}(a)
\end{array}$
\vspace{0.5em}

\noindent
The operations that define the CleverCSV AI assistant follow the same structure as those of datadiff
and are shown in Appendix~\ref{app:details} (see supplemental files). The $\best_\Din$ operation
optimizes the objective function $Q(\Din, e)$ over a set $E_H$, consisting of dialects compatible
with the current constraints $H$. The $\choices_{\Din}(H)$ operation can take
advantage of the consistency score $Q(X, e)$ computed for each dialect under consideration to sort
the suggested constraints.

\parag{Objective function optimization}
The objective function $Q_H$ for the AI assistant does not depend on user interactions and uses
the consistency measure of non-interactive CleverCSV \cite{ccsv}. The measure is
calculated by parsing the input file using a potential dialect and taking the
product of two scores: the
``pattern score'' that captures how regular the structure of the parsed data is (i.e.,~does the
resulting table have the same number of cells in each row?), and the ``type score'' that captures
the proportion of cells that have an identifiable data type. The optimization
involves iterating over each possible dialect allowed by the constraints $H$, and identifying the
one that maximizes the objective function. Further details on the optimization and runtime
performance of CleverCSV can be found in Appendix~\ref{app:performance}.

\parag{Example of using CleverCSV}
While the automatic dialect detection proposed in \cite{ccsv} achieves 97\% accuracy, one type of
failure arises when there are \emph{two} delimiters that result in consistent row lengths and
interpretable cells:

\vspace{0.4em}
\begin{verbatim}
"{""name"":""John"",""age"":""28""}",22:34:00,01:16:40
"{""name"":""Sara"",""age"":""26""}",18:28:02,19:32:37
"{""name"":""Bill"",""age"":""31""}",02:51:34,10:14:58
"{""name"":""Jane"",""age"":""18""}",13:06:36,16:59:47
\end{verbatim}
\vspace{0.4em}

\noindent
A dialect with colon (\texttt{:}) as the column separator maximizes the consistency measure
even though comma~(\texttt{,}) is the correct separator. This happens because splitting the data on
the colon character results in regular row lengths and because the JSON syntax in the first column is
an unknown data type for CleverCSV. The correct dialect receives the second-highest
consistency score and it differs from the chosen dialect only in the delimiter character. This
can be corrected with single interaction. In fact, $\choices_{\Din}(H)$ could automatically
propose the constraint $\ident{fix\_delimiter}(\verb+,+)$ first.


\subsection{ptype: Inferring column types}
\label{sec:aia-ptype}

After parsing data, the next step is often to identify the data types for each column. This
becomes challenging in the presence of missing and anomalous data. The probabilistic type
detection package ptype \cite{ptype} uses a Probabilistic Finite-State Machine model to solve
this problem with an overall accuracy of 93\%, but lower for data types like dates. We recast
ptype as an interactive AI assistant that allows the data analyst to correct errors in those
situations.

\parag{Formal definition of ptype}
For simplicity, we consider input data $\Din$
with just a single column. The expression $e$ represents inferred information for the column
and consists of the inferred column type $\tau$ and sets of values which (conditional on that type)
are deemed missing and anomalous. Human interactions $H$ allow the analyst to constrain the
type ($\tau$), missing values ($u$), and anomalous values ($v$):

\vspace{0.4em}
\noindent
$\quad\begin{array}{lcll}
e &=& (\tau, \{u_1, \ldots, u_k \}, \{ v_1, \ldots, v_l \})  \\[0.25em]
c &=& \ident{not\_type}(\tau) ~|~ \ident{not\_missing}(u) ~|~ \ident{not\_anomaly}(v)\\
\end{array}$
\vspace{0.4em}

\noindent
The $\ident{not\_type}(\tau)$ constraint marks $\tau$ as an incorrect column type, while
$\ident{not\_missing}(u)$ and $\ident{not\_anomaly}(v)$ prevent ptype from treating values
$u, v$ as missing and anomalous.

\parag{Probabilistic model}
The objective function for ptype is derived from a probabilistic model that views
expressions $e$ as parameters of the transformation $f$ and human interactions $H$ as a
meta-parameter that adjusts the likelihood of values in the parameter space. The $Q_H(X, e)$
function is derived from two probability distributions:

\vspace{0.2em}
\begin{itemize}
\item[--] $p_{H}(\Din\,|\,e)$ denotes the likelihood of the input data $\Din$ given an expression $e$,
  which represents a type alongside with missing and anomalous values
\item[--] $p_{H}(e)$ is a distribution over the expressions, representing prior beliefs
    about probabilities of ex\-pressions, i.e.,~types with missing and anomalous values
\end{itemize}
\vspace{0.2em}

\noindent
The probability distributions are written as $p_H$ because, in general, a human interaction can
change the shape of the distribution as well as its parameters. In the case of ptype, human
interactions do not affect the probability distributions, but are used later when selecting the
solution from a distribution over types.

The objective of a probabilistic AI assistant such as ptype is to maximize the posterior
probability distribution of the set of expressions given the data. This is obtained from the
prior distribution over the expressions $p(e)$ and the likelihood model $p(\Din\,|\,e)$
using Bayes' rule:

\vspace{0.5em}
\noindent
$\qquad Q_H(X,e) = p_H(e\,|\,\Din) \propto p_H(\Din\,|\,e) p_H(e)$
\vspace{0.5em}

\noindent
The operation $\best_\Din(H)$ then takes the type with the highest probability
according to $Q_H(X,e)$ that is compatible with the constraints specified by the user.
If there are no constraints, this is the maximum a posteriori (MAP) solution. If the MAP solution
is incorrect, the analyst can choose the $\ident{not\_type}$ constraint to obtain the next most likely value.

\parag{Implementation}
Since non-interactive ptype \cite{ptype} infers the posterior distribution, our interactive
tool only needs to select the most likely solution compatible with the specified constraints.
Interestingly, this is a general approach which can be implemented for any tool based on
a probabilistic framework, regardless of the particular problem it solves.

When generating constraints, $\choices_\Din$ allows the analyst to mark a type as incorrect,
but also to mark values inferred as anomalous or missing as valid. This forces the assistant
to choose the best type that considers them as normal.
The formal model of the logic is given in Appendix~\ref{app:details}.


\subsection{ColNet: Semantic type prediction}
\label{sec:aia-colnet}

Annotating data with semantic information can further assist data analysis. Tools like
OpenRefine~\cite{openrefine} and ColNet~\cite{colnet} automatically annotate tabular data with
semantic types such as \ident{dbo:Company} and \ident{dbo:Person} obtained from a knowledge
graph~\cite{kgs2020} such as DBpedia~\cite{lehmann2015dbpedia}. This may fail when data
contain ambiguous values such as ``Apple'' or ``Virgin'' or values that
do not exist in the knowledge graph (e.g.,~non-famous people \cite{SemTab2019}). We present
an AI assistant based on ColNet (currently under development) that lets the analyst resolve
such errors.

\parag{Formal definition of ColNet}
ColNet is an optimization-based AI assistant, but it has a different structure than datadiff and
CleverCSV. For simplicity, we consider a single-column input $\Din$ formed by a set of values~$v_{i}$.
When inferring the semantic type, ColNet uses a set of samples $S_1, S_2, \ldots, S_n$ which
each contain several individual values from the input data. The sampling method is discussed
in~\cite{colnet}.

The expression produced by the assistant is a single semantic type $\sigma$, to be attached to the
dataset. The analyst can influence the result by specifying a list of constraints. The constraints
$\ident{is\_type}(S, \sigma)$ and $\ident{not\_type}(S, \sigma)$ override the automatically
inferred type for a given sample $S$.

In contrast to the constraints used in ptype, the
constraints used by ColNet override the type of individual samples, rather than the overall type
of a column. A constraint does not fix a type of the column, but merely provides a hint regarding
one of several samples.

\parag{Objective function optimization}
Non-interactive ColNet \cite{colnet} pre-trains a Convolutional Neural Network (CNN) model for each (relevant)
semantic type in the knowledge graph and fine-tunes the model with information from the
column to be annotated. The CNN is then used to rank the possible semantic types obtained by
querying the knowledge graph.

Given a set of samples $\mathbb{S}$ from a given column, non-interactive ColNet predicts
a score $p_S^{\sigma}$ in $[0,1]$ for each sample $S\in\mathbb{S}$ and semantic type $\sigma$.
The score indicates the likelihood that values in $S$ have a type $\sigma$.
ColNet then averages scores over given samples (i.e.,~$p_{\mathbb{S}}^{\sigma} =\frac{1}{|\mathbb{S}|} \sum_{S\in \mathbb{S}} p_S^{\sigma}$)
and chooses the semantic type $\sigma$ with the largest score. In interactive ColNet,
human interactions affect the scoring of samples.
Assuming $p_S^{\sigma}$ is the score given by non-interactive ColNet,
the interactive AI assistant uses $q_{H,S}^{\sigma}$ defined as:

\vspace{0.75em}
$q_{H,S}^{\sigma} =
  \left\{
  \begin{array}{ll}
    1 & \textnormal{when}~\ident{is\_type}(S, \sigma)\in H\\
    0 & \textnormal{when}~\ident{not\_type}(S, \sigma)\in H\\
    p_S^{\sigma} & \textnormal{otherwise}
  \end{array}
  \right.$
\vspace{0.75em}

\noindent
The objective function $Q_H(X, e)$ is defined by the overall score
$q_{H,\mathbb{S}}^{\sigma}$, computed as the average of scores
of individual samples, i.e.,~$q_{H,\mathbb{S}}^{\sigma} =\frac{1}{|\mathbb{S}|} \sum_{S\in \mathbb{S}} q_{H,S}^{\sigma}$.

The $\best_\Din$ operation searches for a semantic type $s$ (from a knowledge graph $G$)
that maximizes the objective function $Q_H$. The constraints offered by $\choices_\Din$ allow the
analyst to mark any of the samples $S\!\in\!\mathbb{S}$ as either having or not having a predicted type
and are generated as follows:

\vspace{0.75em}
$\begin{array}{l}
\choices_\Din(H) =  \\[0.25em]
\quad \{ \ident{is\_type}(S, \sigma),\ident{not\_type}(S, \sigma) ~|~ S\!\in\!\mathbb{S}, \sigma\!\in\!G, p_{S}^{\sigma}\!\geq\!\epsilon \}\\
\end{array}$
\vspace{0.75em}

\noindent
To offer only relevant types, the constraint generation can be limited to types with a
score greater than a threshold $\epsilon$.


\parag{Example of using ColNet}
One of the columns in the broadband quality data (Section~\ref{sec:motivation}) includes company
names Virgin, BT, Sky, and Vodafone. Non-interactive ColNet predicts
the semantic types (with an associated score in parentheses): \ident{dbo:Work} ($0.6$),
\ident{dbo:Company} ($0.5$) and \ident{dbo:Person} ($0.4$).

The correct type \ident{dbo:Company} is not in the top position. This case is complex due to the
use of acronyms (BT) and ambiguous entries (Virgin). In the case of Virgin, the expected entity is \ident{dbr:Virgin\_media},
but ColNet also finds \ident{dbr:Virgin\_of\_the\_Rocks} (a painting
of type \ident{dbo:Work}) and \ident{dbr:Mary,\_mother\_of\_Jesus} (type \ident{dbo:Person}).

To resolve the ambiguity regarding Virgin and obtain the expected semantic type,
the analyst can specify a constraint $\ident{is\_type}(\{\text{``Virgin''}\}, \ident{dbo:Company})$,
which fixes the semantic type for the sample $S\!=\!\{\text{``Virgin''}\}$. This constraint
indirectly decreases the likelihood that the types \ident{dbo:Work} or \ident{dbo:Person} will
be inferred as the best semantic type.


\section{Evaluation}
\label{sec:eval}

The previous section shows that the notion of an AI assistant captures a wide range of practical
semi-automatic data wrangling tools. In this section, we evaluate the specific AI
assistants that were presented. In Section~\ref{sec:eval-qualitative}, we use three scenarios to
compare our tools with the state of the art systems. In Section~\ref{sec:eval-quantitative}, we
quantify how many human interactions are needed to obtain the correct result with AI assistants
in cases where the state of the art automatic~tool~fails.
Performance is discussed in Appendix~\ref{app:performance} (see supplement).

For the evaluation, we use real-world datasets from various sources with manually identified ground
truth (CleverCSV, ptype) or synthetic dataset with ground truth known by construction (datadiff).
A summary of datasets used can be found in Table~\ref{tab:datasets} in the Appendix (see supplemental files).


\subsection{Data wrangling scenarios}
\label{sec:eval-qualitative}

We first consider four real-world data wrangling scenarios based either on a problem from the
literature \cite{nazabal2020aida,ptype,ccsv} or earlier data analyses done by the authors.


\begin{table}
\caption{Comparing Ofcom broadband quality data for 2014 and 2015.}
\label{tab:bb-columns}
\vspace{-0.5em}
\centering
\renewcommand{\arraystretch}{1.1}
\setlength{\tabcolsep}{0.6em}
\begin{tabular}{llll}
  \toprule
  Name ('15) & Col ('15) & Name ('14) & Col ('14)\\
  \midrule
   UL24hrmean & 18 & Upload (Mbit/s)24-hour & 13 \\ 
   Web24hr   & 32 & Web page (ms)24-hour & 28\\
   DL24hrmean & 14 & Download (Mbit/s) 24 hrs & 10 \\ 
   URBAN2     & 10 & Urban/rural & 4\\
   Nation     & 11 & N/A & N/A \\
   Latency24hr & 30 & Latency (ms)24-hour & 24\\
  \bottomrule
\end{tabular}
\vspace{-1em}
\end{table}


\subsubsection{datadiff: Merging Broadband data}
\label{sec:eval-ddiff}

For datadiff, we expand the example from Section~\ref{sec:motivation}. The analyst
wants to analyze the change in broadband quality and needs to merge data for years 2014 and
2015. She selects six columns from 2015 and uses datadiff to find corresponding columns
from 2014. Table~\ref{tab:bb-columns} shows the relevant column names and indices, which have
changed between years. Note that ``Nation'' (a categorical column with values England, Wales,
Scotland) has been added in 2015. The analyst obtains the correct result after three interactions:
\begin{enumerate}
\item datadiff matches \texttt{Nation} with \texttt{LLU}, a categorical
  column with three values (LLU, non-LLU and Cable). The analyst chooses
  ``Don't match LLU and Nation''.
\item datadiff matches \texttt{Nation} with \texttt{Urban.rural}, another categorical
  column with three values. The analyst selects ``Don't match Urban.rural and Nation''.
\item datadiff matches \texttt{Nation} with \texttt{Technology}, yet another categorical
  column with three values. The analyst chooses ``Don't match Technology and Nation'.
\item datadiff correctly identifies that \texttt{Nation} has no corresponding column in 2014
  and generates an \emph{insert} patch to add a new empty column.
\end{enumerate}
\noindent
In all three interactions, the analyst immediately notices that there is one incorrectly
matched column and selects a \ident{nomatch} constraint. In non-interactive datadiff~\cite{datadiff},
the analyst would have to manually edit the initial set of patches (returned as an R object)
or tweak one of the datadiff hyperparameters. Either of those is more complex than choosing
three constraint with informative labels.

In Trifacta~\cite{trifacta}, the same task can be solved by using the ``Add Union'' operation. Here,
the analyst chooses the 2015 dataset, selects the desired 6 columns and then adds the 2014 dataset.
Choosing ``auto align'' invokes a proprietary algorithm that attempts to find matching columns
using column names, column types, and similarity between sampled data.

At the time of writing, the algorithm aligned two of the columns (``UL24hrmean'' and ``Latency24hr'')
and provided no mapping for the remaining four that have to be matched manually using a graphical
user interface. In other words, Trifacta is less successful in guessing the initial
matching but, more importantly, it also only implements the \emph{onetime interaction} model where
analyst invokes the automatic tool once, but then has to correct all errors manually.


\subsubsection{CleverCSV: Parsing large and messy CSV files}

Dialect detection can seem a trivial task, but large CSV files can hide problems that are
difficult to detect manually. We consider two scenarios: one where CleverCSV infers the dialect
correctly and one where a single human interaction is needed.

First, consider the Internet Movie Database file\footnote{From: \url{https://www.kaggle.com/orgesleka/imdbmovies}.},
which contains descriptive statistics for 14,762 movies. A few rows and columns from the file look as follows:

\vspace{0.25em}
\begin{verbatim}[numbers=left,xleftmargin=6mm]
fn,title,imdbRating
titles01/tt0015864,Goldrausch (1925),8.3
titles01/tt0017136,Metropolis (1927),8.4
titles01/tt0017925,Der General (1926),8.3
titles02/tt0080388,Atlantic City\,USA (1980),7.4
\end{verbatim}
\vspace{0.25em}

\noindent
In this case, CleverCSV infers the correct dialect fully automatically.
The standard R and Python functions fail to identify the escape character (\texttt{\textbackslash})
which is used for movies with a comma in the title (line 5) and load 15,190 and 13,928 rows, respectively.
Trifacta \cite{trifacta} also does not correctly handle the escape character. It assumes the file
has three columns due to the first row and silently merges the additional data into the last column.
OpenRefine \cite{openrefine} instead adds a fourth column due to the fact that the last row contains
three delimiters. Such failures can be very time consuming to address and neither Trifacta nor
OpenRefine provide straightforward mechanisms to mitigate this problem.

In cases when CleverCSV does not automatically detect the correct dialect, the AI assistant
shows a preview of the parsed CSV file to the analyst, who can steer CleverCSV in the right
direction. The following shows a few lines of a CSV file that contains filenames and
RGB color codes\footnote{Available at: \url{https://github.com/victordiaz/color-art-bits-}.}:

\vspace{0.25em}
\begin{verbatim}[numbers=left,xleftmargin=6mm]
1894_0.jpg   51,47,45   87,88,86    110,112,110
1895_0.jpg   37,25,24   87,59,47    105,88,88
1895_1.jpg   48,34,46   80,51,58    98,80,88
1901_0.jpg   45,46,55   100,96,91   115,139,129
1901_1.jpg   45,46,48   71,66,61    98,97,94
\end{verbatim}
\vspace{0.25em}

\noindent
For this file, CleverCSV predicts the comma as the delimiter even though the tab character is used.
The analyst notices the issue easily thanks to the provided preview and can fix the parsing through
a single interaction: by choosing the $\ident{fix\_delimiter}(\texttt{\textbackslash t})$ constraint
to set the correct delimiter.

For the same file, OpenRefine chooses underscore as the delimiter, whereas Trifacta
uses the comma character. While in this case the user can select the correct delimiter in OpenRefine,
this is not the case in Trifacta, where additional manual interaction is needed to get the data to
a usable state.


\subsubsection{ptype: Annotating the Cylinder Bands dataset}

For the type inference task, we consider the Cylinder Bands dataset from the UCI repository \cite{uci}.
The file contains data on process defects known as ``cylinder bands'' in rotogravure printing.
When analyzing the file, ptype fails to correctly identify the type for some columns of this
dataset.

For example, the ``ESA Amperage'' column contains mostly the \texttt{0} value (480 out of 540 entries)
and a small number of other values (\texttt{0.5}, \texttt{4}, \texttt{6}, \texttt{?}). The initial
type offered by ptype is Boolean with \texttt{0.5}, \texttt{4} and \texttt{6} incorrectly treated as
anomalies and \texttt{?} correctly identified as missing data. This is perhaps unsurprising given
the dominance of \texttt{0} values.

The analyst can obtain the correct type through a single interaction, by choosing ``ESA
Amperage is not Boolean'', which adds the $\ident{not\_type}(\texttt{Boolean})$ constraint.
The assistant then returns the correct, second most likely, data type \texttt{Float} with
no anomalies and \texttt{?} as the missing data indicator.

State of the art tools face similar issues. Trifacta labels the ``ESA Amperage'' column with the
integer type rather than float. It considers \texttt{4} and \texttt{6} valid values, but
\texttt{0.5} and \texttt{?} are treated as
\emph{mismatched values}. The analyst needs to change the assigned type to float through the
user interface, by clicking on the integer sign and then selecting the float type. This interaction
is specific to type inference in Trifacta and requires familiarity with the graphical user interface.

OpenRefine does not directly address column-type inference. Instead, it separately infers the type
for each entry. It correctly identifies the data type for the entries of ``ESA Amperage'' as it uses
the numeric label rather than separate float and integer types. However, user interaction is
required for many other columns in the same dataset that are labeled correctly by both ptype and
Trifacta. For example, the ``grain screened'' column represents a Boolean with values
\texttt{yes} and \texttt{no}. Here, ptype and Trifacta correctly infer the type as \texttt{Boolean},
whereas OpenRefine treats the values as text. Changing the assigned type to
\texttt{Boolean} converts both \texttt{yes} and \texttt{no} to \texttt{false}. To get the correct
types, the analyst first needs to replace all values of \texttt{yes} with \texttt{true}.


\subsection{Empirical evaluation}
\label{sec:eval-quantitative}
For optimization-based AI assistants, we can evaluate how many interactions are needed to
arrive at the correct result. As each assistant solves a different task, we need to use a
different dataset for each. Table~\ref{tab:eval-interactions} shows the results; for each
AI assistant, we show the fraction of cases that requires a specific number of interactions.
The ``Average'' column shows the average over the cases where \emph{some} human interaction
is required. The datasets used are discussed below.


\begin{table}
\caption{Interactions required to solve a wrangling task for each AI assistant.}
\label{tab:eval-interactions}
\vspace{-0.5em}
\centering
\renewcommand{\arraystretch}{1.1}
\setlength{\tabcolsep}{0.6em}
\begin{tabular}{lccccccl}
\toprule
  AI assistant (dataset)  & \multicolumn{5}{c}{Number of interactions} & Average\\
                & 0 & 1 & 2 & 3 & 4+ &  \\
  \midrule
   datadiff (UCI)  & 0.52 & 0.20 & 0.12 & 0.00 & 0.18 & 3.25\\
   datadiff (without Iris) & 0.63 & 0.22 & 0.15 & 0.00 & 0.00 & 1.40\\
   CleverCSV (GitHub) & 0.20 & 0.70 & 0.05 & 0.04 & 0.00 & 1.17\\
   ptype     (Various) & 0.33 & 0.51 & 0.16 & 0.00 & 0.00 & 1.24\\
  \bottomrule
\end{tabular}
\vspace{-1em}
\end{table}


\parag{datadiff}
Following the original datadiff evaluation~\cite{datadiff} we use a synthetic dataset
obtained by corrupting five datasets from the UCI repository~\cite{uci} (Abalone,
Adult, Bank, Car, Iris). To corrupt a file, we randomly reorder columns and apply two
other randomly chosen corruptions.

The corruptions include inserting a numerical column
(with values from a uniform distribution $U(0,1)$), inserting a categorical column
(with two evenly distributed values), deleting a random column, recoding a
categorical column and applying a linear transformation (with $a$ from $U(-0.5,0.5)$
and $b$ from $U(-2\overline{v},2\overline{v})$ where $\overline{v}$ is the mean of the
values in the column). We apply the corruption to a randomly
selected half of the data and attempt to reconcile the two halves using datadiff.
When datadiff does not produce the expected result, we repeatedly add \ident{nomatch} constraints
to prevent incorrect matchings inferred by datadiff.

Our corruptions and dataset are more challenging than those used previously~\cite{datadiff}.
We note that datadiff performs poorly on one of the five datasets (Iris), so the table shows
results for all five datasets as well as for the remaining four (without Iris). Datadiff
requires no human interaction in 52\% and 63\% cases, respectively. Our evaluation models the case where the
analyst can easily spot an error and inform the assistant, but the number of interactions could be
reduced further by choice of the explicit \ident{match} constraint.


\parag{CleverCSV}
To evaluate CleverCSV, we revisit the failure cases of the non-interactive CleverCSV
\cite{ccsv} and count interactions needed to find the correct
dialect. We apply the assistant on 255 files from a corpus of CSV files extracted
from GitHub where the dialect was detected incorrectly in~\cite{ccsv}. We focus on this selection
as the 97\% of cases where CleverCSV detects the dialect correctly
are not relevant here. Since the dialect considered for the CSV file consists of three components,
the maximum number of interactions is three.

For 20\% of the 255 files no interaction is needed to find the correct dialect. This can be
attributed to improvements in
CleverCSV since publication of \cite{ccsv}. For the majority of files (70\%) a single interaction
was needed, with an average of only 1.17. This illustrates that as the human provides a constraint
to the AI assistant, the limits on the search space allow CleverCSV to quickly arrive at the correct answer.

\parag{ptype}
To evaluate the ptype AI assistant, we consider 43 (out of 610) data columns where the types were not
inferred correctly by the non-interactive ptype~\cite{ptype}, using a corpus obtained from various
sources including the UCI repository~\cite{uci} and open government data sources. To guide the
assistant, we iteratively add the \ident{not\_type} constraint.

Although ptype recognizes 11 data types, we focus on 5 primitive types (Boolean, integer,
floating-point number, date, string) to allow comparison with other tools. Even with 5 types,
identifying them correctly by hand remains difficult, because ptype also detects anomalous and
missing values, which may not be easy to notice for a human analyst.

Of the 43 data columns, no interaction is needed for 33\% of cases. As
with CleverCSV, this is due to recent improvements in ptype. A single interaction was
needed for a majority of files (51\%). In those cases, the assistant arrives at the correct answer
by choosing the second most likely type. The remaining columns require two interactions, resulting in
an average of 1.24. Note that rejecting the offered type is a simpler interaction than directly
selecting a type, which would reduce the maximum number of interactions to one.

\parag{Summary}
The quantitative evaluation of three AI assistants demonstrates that many data
wrangling tasks can be solved much more efficiently by creating opportunities where the human
analyst can nudge the tool in the right direction. This approach obviates the need for tedious
data manipulation in spreadsheet applications or case-specific wrangling scripts, and is
significantly easier to implement than fully automatic tools that need to cover numerous edge cases.


\section{Related work}
\label{sec:related}

The problem of data wrangling has been studied by both practitioners~\cite{crowdflower2016data,kaggle2017state,challenges}
and academics~\cite{kandel2011research,cleaningstudy,gorinova2016end,casestudies}. These studies
repeatedly mention the problems that motivated our work. We believe that \emph{interactivity}, and
\emph{uniformity} are crucial. Interactivity allows incorporating crucial human insights, and a
common structure makes it possible for the analyst to easily access a wide range of tools.

\parag{Programming and analytic systems}
Data wrangling is often done programmatically in the R and Python languages,
using libraries such as Tidyverse and Pandas \cite{tidyverse,pandas} in notebook systems
like RStudio and Jupyter. Our AI assistants are available for the Jupyter platform through
the Wrattler extension \cite{wrattler}, which enables polyglot programming.

Spreadsheets and business intelligence tools such as Tableau and Power BI
provide a complex set of features for data analytics, often used through a
complex graphical user interface, while more focused data wrangling tools like Trifacta
and OpenRefine \cite{trifacta,openrefine} provide similar environments focused
on data cleaning. As discussed in Section~\ref{sec:eval}, those tools address many of the
specific problems addressed by AI assistants, but lack uniformity and rarely implement the
powerful \emph{iterative} interaction model.

\parag{Data wrangling and repair tools}
A number of tools attempt to solve a specific data wrangling problem automatically, including the
tools extended in this paper \cite{datadiff,ptype,ccsv,colnet} as well as tools for data imputation
\cite{hivae}, deduplication \cite{duplicate}, and parsing \cite{learnpads}.
These tools often achieve a high accuracy, but they lack an easy-to-use mechanism for incorporating
critical human insights in cases where the automatic answer is incorrect. Automatic tools can
also be guided by a manually written domain-specific data model, as in PClean \cite{lew2021pclean}.

A few systems utilize the flexible \emph{iterative} interaction pattern to suggest
possible data transformations using machine learning. Proactive wrangling \cite{guo2011proactive}
suggests data transformations to improve data structure based on a metric and offers those to the
user. In Predictive Interaction \cite{heer2015predictive}, inputs provided by the analyst are used
to generate a ranked list of predictions from which the analyst can choose or, alternatively,
provide further inputs. This is similar to how AI assistants work, but the way of specifying
feedback is domain-specific, e.g.,~highlighting substrings in textual~data.

The problem of incorporating user input has been extensively studied in data repair tools for
databases \cite{dance,potter}. Tools for enforcing functional database dependencies
\cite{repair,uguide,fan2008conditional} work by asking analysts questions about the data
and using the answers to improve the model used for data repair.  Such tools could be recast as
AI assistants; they complement our examples in that they focus on working with databases whereas
our focus is on less structured data.

\parag{Programming language approaches}
Numerous programmatic tools offer a small domain-specific language for a particular data wrangling
task such as statistical analysis or data visualization \cite{tea,vegalite}. A small domain-specific
language is also at the core of semi-automatic tools such as LearnPADS and Predictive Interaction
\cite{learnpads,heer2015predictive}. AI assistants follow the same approach in that the expressions
of individual AI assistants form small languages that are easy to understand.

AI assistants can also be seen as a form of code completion. This typically focuses on offering
available operations, possibly using machine learning to rank the recommendations \cite{mlcomplete}.
Type providers \cite{inforich,dotdriven} are closer to our approach in that the recommendations are
generated programmatically, similar to our $\choices$ operation.

\parag{Human-computer interaction}
Two interaction techniques used in data wrangling tools are direct manipulation and
programming-by-example. In the former, a program is specified by directly interacting with the
output. This has been used for data analysis and querying \cite{control,dynamicq,visage,dataplay},
as well as data wrangling \cite{potter}. In the latter, the user gives examples of
desired results, for example, to specify data transformations in spreadsheets
\cite{excelbyexample}. This results in an \emph{iterative} interaction mechanism, but one where
the analyst needs to specify more complex inputs as opposed to just choosing from a list of options.
Novel human-interaction techniques for data wrangling also include touch-based editing
\cite{vizdom}, natural language \cite{eviza}, and conversational agents~\cite{iris}.

\parag{Human in the loop}
Our work contributes to the emerging field of human-in-the-loop data analytics
\cite{li2017human,doan2018human}. AI assistants particularly implement the ``efficient correction''
pattern \cite{guidelines}. We focus on supporting an individual analyst, but
a range of systems involve multiple users in addressing data wrangling problems.
In \cite{endtoend}, data cleaning problems are solved by assigning the tasks that cannot be
automated to human ``detectors'' and ``repairers'' and several data cleaning tools rely
on crowdsourcing \cite{nadeef,katara,crowder}.

\section{Future work}
\label{sec:future}

Allowing AI assistants to accept richer user inputs would let us support programming-by-example.
Programming-by-example can be seen as ranking programs in an underlying DSL that are consistent
with a given set of training examples \cite{gulwani2016programming},
fitting well with our optimization-based AI assistant structure. Alternatively, focusing on
probabilistic AI assistants would let the system leverage additional information, such as the
distribution of possible cleaning scripts, allowing users to choose a desired solution more effectively.
The usability of AI assistants could also be improved by offering possible choices in a
more structured way than as a flat list.

The AI assistants presented in this paper solve individual data wrangling problems, but a
typical data wrangling workflow involves a combination of tools. An interesting direction for future
work is to recommend the entire data wrangling workflow, composed of multiple AI assistant invocations.
This could be done by repeatedly predicting the next step as in \cite{heer2015predictive,guo2011proactive}.
Closer interaction between AI assistants could also lead to better results. For example, the
consistency measure used by CleverCSV could incorporate information obtained from ptype or ColNet,
while datadiff could prefer matching columns with the same data type, as inferred by ptype or ColNet.

Finally, AI assistants do not currently learn from past user interaction.
Using the interactions with human analysts to improve the models underlying the AI assistants
as well as learning the ways in which AI assistants are composed could provide valuable
information for improving the accuracy of the inference done by the assistants.

\section{Conclusion}
\label{sec:sum}

Data wrangling is notoriously tedious and hard to automate. It has eluded the recent rise
of AI because large datasets hide corner cases that require human insight.
We have introduced the notion of AI assistants, which captures the structure
of semi-automatic, interactive tools for data wrangling. We showed how the definition captures
common types of tools and makes them easy to use from notebook systems. We developed
four concrete AI assistants that are flexible interactive versions of existing
non-interactive tools.

This paper makes two claims. First, we argue that the structure of AI assistants is a suitable
abstraction for interactive data wrangling tools. Second, we argue that our interactive AI
assistants are more practical than fully automatic tools. To support the first claim,
we present AI assistants that cover a wide range of data wrangling tasks including
parsing, merging mismatched datasets, type inference, and the inference of
semantic information. To support our second claim, we discuss three real-world case studies
where a fully automatic tool does not give the desired result, together with an empirical
evaluation that showed that users can typically solve a wrangling task with 1-3 simple
interactions.

While we cannot hope to reduce to zero the 80\% of the time that data analysts spend on data wrangling
solely with what we have described above, we believe that our framework provides the right pathway.
A growing ecosystem of interactive unified AI assistants would allow data
analysts to fully leverage recent AI advances for the most tedious and time-consuming aspect of
their job and pave the way for more equitable access to data science and machine learning.

\section*{Acknowledgments}
The work was supported in part by EPSRC grant EP/N510129/1 to The Alan Turing Institute.
TP and CW would like to acknowledge the funding provided by the UK Government's Defence \& Security
Programme in support of The Alan Turing Institute. The work of EJR was also supported in part by
the SIRIUS Centre for Scalable Data Access (Research Council of Norway, project 237889) and TC
was supported by a PhD studentship from The Alan Turing Institute (EPSRC grant TU/C/000018).

The work on AI assistants would not be possible without work on Wrattler by May Yong
and Nick Barlow. We benefited from discussions with colleagues in The Alan Turing Institute,
especially Charles Sutton and James Geddes. We would also like to highlight the contribution of
Jiaoyan Chen, the researcher behind the non-interactive ColNet. Last but not least, we
thank the anonymous reviewers for their comments that have helped improve the paper.

For the purpose of open access, the authors have applied a Creative Commons Attribution (CC BY) licence to any Author Accepted Manuscript version arising from this.

\bibliographystyle{ieeetr}
\bibliography{paper}

\begin{IEEEbiography}[{\includegraphics[width=1in,height=1.25in,clip,keepaspectratio]{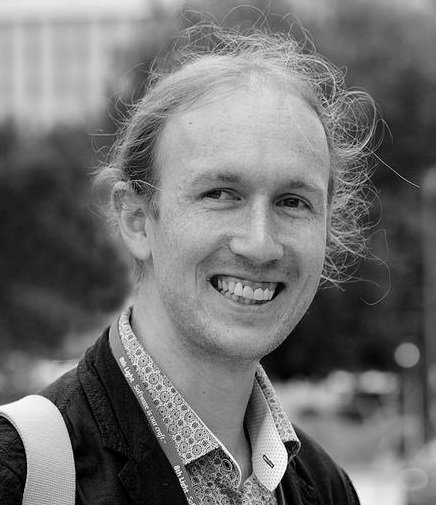}}]{Tomas Petricek}
is a Lecturer in School of Computing at University of Kent, UK. His research focuses on making programming
easier, trustworthy and more accessible. Previously, he worked on tools for data science in The
Alan Turing Institute and functional programming language F\# in Microsoft Research.
He holds PhD from University of Cambridge where he developed theory of context-aware programming
languages.
\end{IEEEbiography}

\begin{IEEEbiography}[{\includegraphics[width=1in,height=1.25in,clip,keepaspectratio]{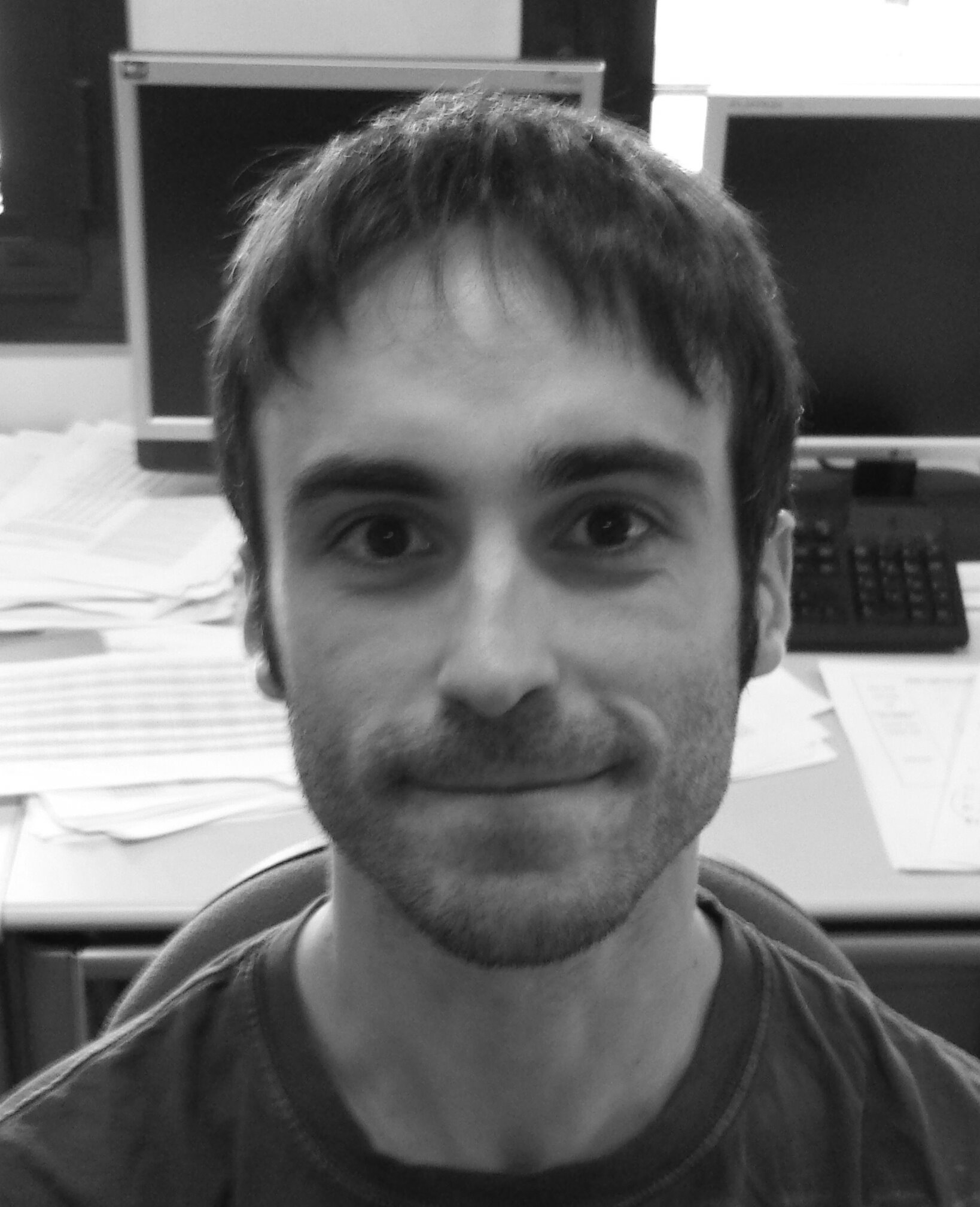}}]{Alfredo~Naz{\'{a}}bal}
is an applied scientist working in the Amazon Development Center in Edinburgh. Before that he was a
Postdoctoral researcher in The Alan Turing Institute in London where he worked developing deep
generative models for data analytics problems. He obtained his PhD in the University Carlos III of
Madrid, where he developed and applied Machine Learning algorithms for Human Activity Recognition.
His main research interests include deep generative models, unsupervised learning, heterogeneous
data preprocessing and recommender systems.
\end{IEEEbiography}

\begin{IEEEbiography}[%
	{\includegraphics[width=1in,height=1.25in,clip,keepaspectratio]{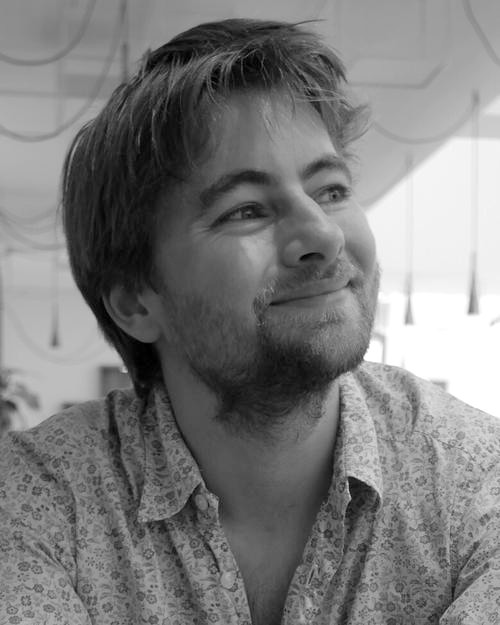}}%
	]{Gerrit~J.J.~van~den~Burg}
	is an applied scientist at Amazon. Previously, he was a postdoctoral
	researcher at The Alan Turing Institute in London, UK, where he worked
	on automating the manual parts of data science using machine learning.
	He obtained a PhD from the Erasmus University Rotterdam in The
	Netherlands, during which he focused on developing algorithms for
	multiclass classification and sparse regularization.
\end{IEEEbiography}

\begin{IEEEbiography}[%
        {\includegraphics[width=1in,height=1.25in,clip,keepaspectratio]{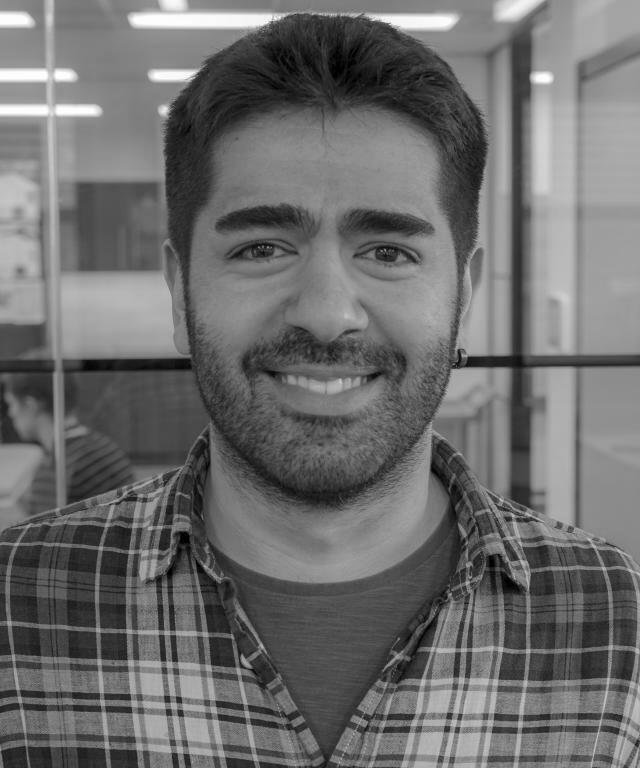}}%
        ]{Taha~Ceritli}
is a postdoctoral research assistant at the University of Oxford. His current research focus is
machine learning for healthcare. He previously received his PhD from the University of Edinburgh
on probabilistic data type inference, which was carried out at and supported by The Alan Turing Institute.

\end{IEEEbiography}

\begin{IEEEbiography}[%
        {\includegraphics[width=1in,height=1.25in,clip,keepaspectratio]{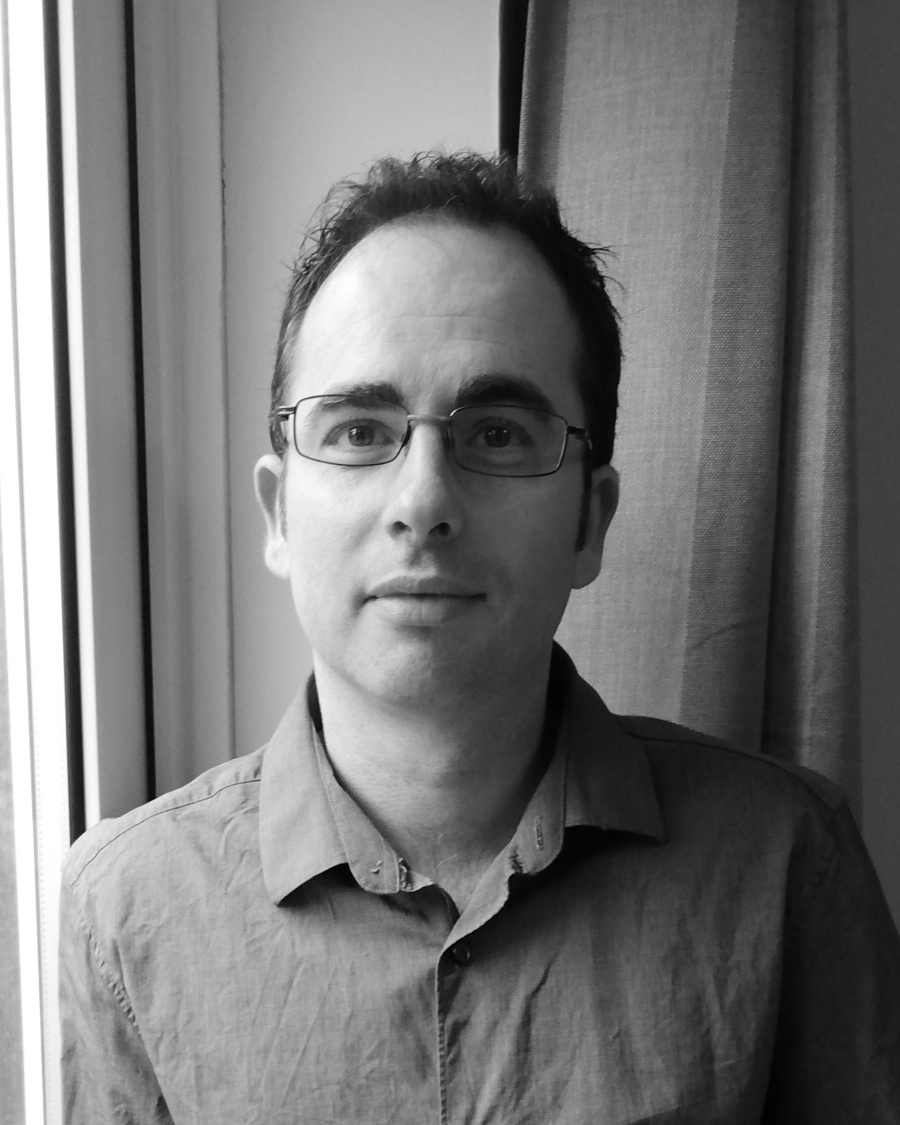}}%
        ]{Ernesto~Jim\'enez-Ruiz}

	is a Lecturer in Artificial Intelligence at City, University of London (UK), and a researcher at the University of Oslo (Norway). He previously held a Senior Research Associate position at The Alan Turing Institute in London (UK) and a Research Assistant position at the University of Oxford (UK). His research interests focus on the application of Semantic Web technology to Data Science workflows and the combination of Knowledge Representation and Machine Learning techniques.
\end{IEEEbiography}

\begin{IEEEbiography}[
        {\includegraphics[width=1in,height=1.25in,clip,keepaspectratio]{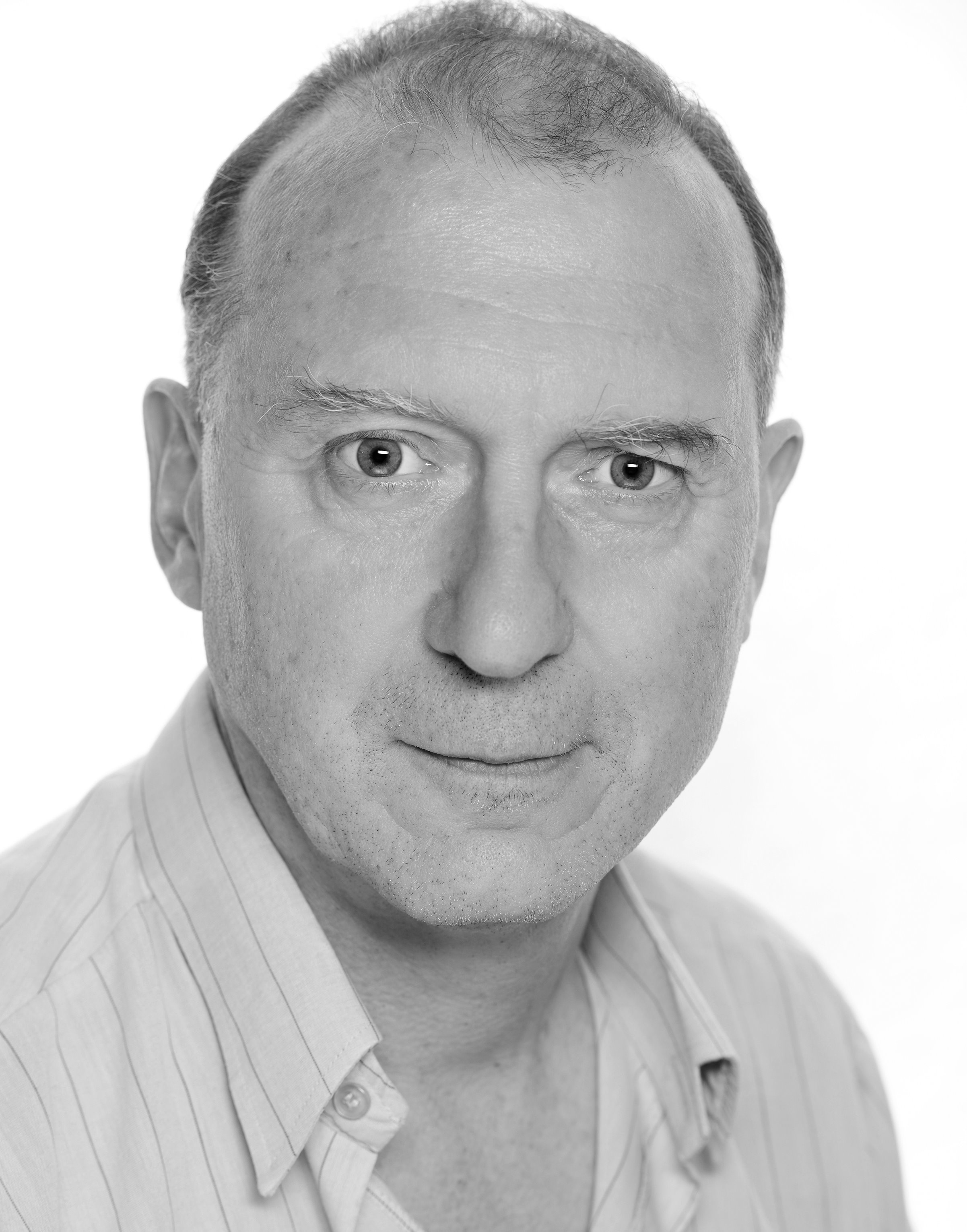}}%
        ]{Chris~Williams}
  is Professor of Machine Learning in the School of Informatics, University of Edinburgh.  His main areas of research are in visual object recognition and image understanding, models for understanding time-series, AI for data analytics, unsupervised learning, and Gaussian processes.  He obtained his MSc (1990) and PhD
  (1994) at the University of Toronto, under the supervision of Geoffrey Hinton.  He was at Aston University 1994-1998, and has been at the University of Edinburgh since 1998.
\end{IEEEbiography}

\newpage

\appendices

\section{Formal definitions}
\label{app:details}

This appendix provides further details of the formal models of the AI assistants discussed in the
main text. The complete descriptions provided here enhance the reproducibility of our work and
make it possible to reimplement AI assistant as presented and evaluated in this paper.
A summary of symbols used in the formalization can be found in Table~\ref{tab:symbols}.

\parag{datadiff}
The definition of the datadiff AI assistant given in Section~\ref{sec:aia-ddiff} shows
patches, constraints, and the $\best_\Din$ operation. It omits the $\choices_\Din$ operation
and the \ident{valid} predicate which identifies patches that are valid for a given set of
human interactions $H$. The operations of the datadiff assistant,
including the \ident{valid} predicate, are defined as:
\begin{equation*}
\begin{array}{l}
\best_\Din(H) = \argmax_{e \in E_H} Q(\Din, e)~\textnormal{where}\\[0.25em]
\quad E_H = \{ (P_1, \ldots, P_k) \in E ~|~ \forall i\in 1\ldots k.\,\ident{valid}_H(P_i) \}
\\[0.75em]
\ident{valid}_H(P)~\textit{such that}\\[0.25em]
\quad \ident{valid}_H(\ident{permute}(\pi)) ~\mathit{iff}~ \\[0.25em]
\quad \quad (\forall \ident{match}(i,j)\in H.~(i,j)\in \pi) ~\wedge~\\[0.25em]
\quad \quad (\forall \ident{nomatch}(i,j)\in H.~(i,j)\notin \pi) \\[0.25em]
\quad \ident{valid}_H(\ident{recode}(i, [\ldots])) ~\mathit{iff}~ \ident{notransform}(i)\notin H\\[0.25em]
\quad \ident{valid}_H(\ident{linear}(i, a, b)) ~\mathit{iff}~ \ident{notransform}(i)\notin H\\[0.25em]
\quad \ident{valid}_H(\ident{delete}(i)) \\[0.25em]
\quad \ident{valid}_H(\ident{insert}(i, d))
\\[0.75em]
\choices_\Din(H) = \\[0.25em]
\quad \{ H \cup \{ \ident{notransform}(i) \}~|~\forall i. \ident{recode}(i, [\ldots])\in e \} ~\cup\\[0.25em]
\quad \{ H \cup \{ \ident{notransform}(i) \}~|~\forall i. \ident{linear}(i, a, b)\in e \} ~\cup\\[0.25em]
\quad \{ H \cup \{ \ident{nomatch}(i, j) \}~|~ \ident{permute}(\pi) \in e, \forall i,j. (i,j)\in \pi \} ~\cup\\[0.25em]
\quad \{ H \cup \{ \ident{match}(i, j) \}~|~ \ident{permute}(\pi) \in e, \forall i,j. (i,j)\notin \pi \}\\[0.25em]
\quad \textnormal{where}~e = \best_\Din(H)
\end{array}
\end{equation*}

\parag{ptype}
In the ptype AI assistant, $\best_\Din$ is obtained by taking the
maximum a posteriori of the posterior probability distribution of
the set of expressions determined by the past human interactions.
As in the case of datadiff and CleverCSV, this is defined using the
\ident{valid} predicate. The $\choices_\Din$ operation allows the analyst
to reject an inferred type and mark an inferred missing or anomalous value as non-missing or
non-anomalous. More formally, the operations of ptype are defined as follows:
\begin{equation*}
\begin{array}{l}
\best_\Din(H) = \argmax_{e \in E_H} p_H(\Din|e)\,p_H(e) ~\textnormal{where}\\[0.25em]
\quad p_H(\Din|e) = p(\Din|e)~\textnormal{and}~p_H(e)=p(e)\\[0.25em]
\quad E_H = \{e \in E \mid \ident{valid}_H(e)\}
\\[0.75em]
\ident{valid}_H(\tau, V_m, V_a)~\textit{iff}\\[0.25em]
\quad (\ident{not\_type}(\tau')\in H\implies \tau'\neq\tau)~\vee\\[0.25em]
\quad (\ident{not\_missing}(v)\in H\implies v\notin V_m)~\vee\\[0.25em]
\quad (\ident{not\_anomaly}(v)\in H\implies v\notin V_a)
\\[0.75em]
\choices_\Din(H) = \{ H \cup \{ \ident{not\_type}(t) \} \} ~\cup\\[0.25em]
\qquad \{ H \cup \{ \ident{not\_missing}(v_j) \} ~|~ j \in J \}~\cup\\[0.25em]
\qquad \{ H \cup \{ \ident{not\_anomaly}(w_k) \} ~|~ k \in K \}\\[0.25em]
\quad \textit{where} \best_\Din(H) = \\[0.25em]
\quad \quad (\ident{type}(t), \ident{missing}\{v_j \}_{j\in J}, \ident{anomaly}\{ w_k \}_{k\in K})
\end{array}
\end{equation*}

\parag{CleverCSV}
The definition of the CleverCSV AI assistant closely follows the example of datadiff. The
$\best_\Din$ operation uses the pattern of the optimization-based AI
assistants. The $\choices_\Din$ operation allows the analyst to reject a
component of a currently inferred dialect or to explicitly choose a specific
character for a dialect component. As before, the \ident{valid} predicate determines what
is a valid dialect given past human interactions. Formally:
\begin{equation*}
\begin{array}{l}
\best_\Din(H) = \argmax_{e \in E_H} Q(\Din, e)~\textnormal{where}\\[0.1em]
\quad E_H = \{ e \in E ~|~ \ident{valid}_H(e) \}
\\[0.75em]
\ident{valid}_H(\ident{is\_delimiter}(d), \ident{is\_quote}(q), \ident{is\_escape}(a))~\textit{iff}\\[0.25em]
\quad (\ident{fix\_delimiter}(d')\in H \implies d'=d) ~\vee \\[0.25em]
\quad (\ident{fix\_quote}(q')\in H \implies q'=q) ~\vee \\[0.25em]
\quad (\ident{fix\_escape}(a')\in H \implies a'=a) ~\vee \\[0.25em]
\quad (\ident{block\_delimiter}(d')\in H \implies d'\neq d) ~\vee \\[0.25em]
\quad (\ident{block\_quote}(q')\in H \implies q'\neq q) ~\vee \\[0.25em]
\quad (\ident{block\_escape}(a')\in H \implies a'\neq a)
\\[0.75em]
\choices_\Din(H) = \\[0.25em]
\qquad \{ H \cup \{ \ident{not\_delimiter}(c_d) \}, H \cup \{ \ident{not\_quote}(c_q) \}, \\[0.25em]
\qquad \,\; H \cup \{ \ident{not\_escape}(c_e) \} ~\}~\cup\\[0.25em]
\qquad \{ H \cup \{ \ident{delimiter}(c) \}~|~\forall c. C \} ~\cup\\[0.25em]
\qquad \{ H \cup \{ \ident{quote}(c) \}~|~\forall c. C \} ~\cup~ \{ H \cup \{ \ident{escape}(c) \}~|~\forall c. C \} \\[0.25em]
\quad \textnormal{where}~(\ident{delimiter}(c_d), \ident{quote}(c_q),
\ident{escape}(c_e)) = \best_\Din(H).
\end{array}
\end{equation*}

\parag{ColNet}
The definition of ColNet differs from the other examples in that human interactions affect
the objective function $Q_H(\Din,e)$ rather than the set of possible expressions $E_H$.
The definition of the objective function is discussed in Section~\ref{sec:aia-colnet}. The
following provides a full definition of both of the operations of the AI assistant
for completeness:

\begin{equation*}
\begin{array}{l}
\begin{array}{lcl}
  c &=& \ident{not\_type}(S, \sigma) ~|~ \ident{is\_type}(S, \sigma)\\[0.5em]
  p_S^{\sigma} &=& \textnormal{As defined in non-interactive ColNet}\\[0.5em]
  q_{H,S}^{\sigma} &=&
  \left\{
  \begin{array}{ll}
    1 & \textnormal{when}~\ident{is\_type}(S, \sigma)\in H\\
    0 & \textnormal{when}~\ident{not\_type}(S, \sigma)\in H\\
    p_S^{\sigma} & \textnormal{otherwise}
  \end{array}
  \right.
  \\[-0.25em]
  q_{H,\mathbb{S}}^{\sigma} &=&\frac{\displaystyle 1}{\displaystyle |\mathbb{S}|}
    \mathlarger{\sum}_{S\in \mathbb{S}} q_{H,S}^{\sigma}\\
  ~\\[-0.5em]
\end{array}\\
\begin{array}{l}
\best_\Din(H) = \argmax_{\sigma \in G}  Q_H(\Din, \sigma)~\textnormal{where}\\[0.25em]
\quad Q_H(\Din, \sigma) = q_{H,\mathbb{S}}^{\sigma} \\
\\[-0.5em]
\choices_\Din(H) =  \\[0.25em]
\quad \{ \ident{is\_type}(S, \sigma),\ident{not\_type}(S, \sigma) ~|~ S\!\in\!\mathbb{S}, \sigma\!\in\!G, p_{S}^{\sigma}\!\geq\!\epsilon \}\\
\end{array}
\end{array}
\end{equation*}

\section{Performance considerations}
\label{app:performance}

In general, interactive AI assistants obtained by adapting an existing non-interactive tool
(datadiff, ptype, CleverCSV, ColNet) invoke the optimization logic of the non-interactive tool,
with small modifications, each time the user interacts with the assistant. The performance
is thus comparable to the performance of the base tools \cite{datadiff,ccsv,ptype,colnet}.
In some cases, however, the interactive AI assistant can reuse previously computed results
and perform more efficiently.

In this section, we briefly discuss performance in typical
real-world scenarios as well as algorithmic complexity of the optimization and possible
performance improvements for the four AI assistants discussed in the paper.

\parag{datadiff}
The datadiff AI assistant uses the algorithm from non-interactive datadiff \cite{datadiff}.
This works in two phases. In the first phase, the algorithm determines
a cost matrix $C_{ij}$ by finding the best patch between each pair of columns. In the second
phase, the algorithm uses the Hungarian algorithm to find the best bipartite matching.
The interactive AI assistant requires two modifications. First, after computing $C_{ij}$,
we set $C_{ij}$ to $0$ for each $\ident{match}(i,j)$ constraint and to $+\infty$ for
each $\ident{nomatch}(i,j)$ constraint. Second, we modify the logic for finding the best patch
to not use a recoding or linear transformation when $\ident{norecode}$ is specified.

The algorithmic complexity of the first phase is $O(n^2)$ in terms of the number of columns $n$,
while the algorithmic complexity of the second phase is $O(n^3)$. For real-world datasets,
however, most of the time is spent in the first phase, generating and evaluating possible
pairwise patches. Reconciling the full broadband quality dataset for 2014 (31 columns) and 2015
(71 columns) takes 35 seconds on a recent computer.\footnote{Laptop with Intel Core i7-1185G7
processor, 15GB RAM, running inside Docker container on Windows 11 OS.} Reconciling the full
2014 (31 columns) with filtered 2015 (6 columns) dataset, as done in the case study in
Section~\ref{sec:eval-qualitative}, takes 5 seconds.

This makes the current implementation useable for smaller datasets. There are two ways in which
the performance could easily be improved. First, the cost matrix (phase one) could be determined
on the first run and then cached. This does not change between runs and would significantly
improve the performance on subsequent interactions. Second, the cost matrix could be determined
based on a sample of the full dataset, possibly improving the initial cost matrix in background
after offering the first recommendation.

\parag{ptype}
The ptype AI assistant computes the posterior probabilities over data types based on non-interactive ptype \cite{ptype} and updates the initially assigned types according to the user feedback. This is achieved by storing the posterior probability distributions rather than re-running non-interactive ptype. Thus, assuming that the number of known column types and therefore the maximum number of user corrections is constant, the complexity of the ptype AI assistant becomes identical to the complexity of non-interactive ptype.

The computational bottleneck in type inference via ptype is the calculation of probability distribution assigned for a data column $\textbf{x}$ by the $k$th PFSM denoted by $p(\textbf{x}|t = k)$. This calculation is carried out by taking into account only unique data entries, for efficiency. Denoting the $u$th unique data value by $x_u$, the computation of $p(x_u |t = k)$ is done via the PFSM Forward algorithm. This has complexity $O(M_k^2 L)$, where $M_k$ is the number
of hidden states in the $k$th PFSM, and $L$ is the maximum length of the data entries. Therefore, the overall complexity of the inference becomes $O(U K M^2 L)$, where $U$ is the number of unique data entries, $K$ is the number of types, and $M$ is the maximum number of hidden states in the PFSMs.

Notice that the complexity depends on data through $U$ and $L$, and does not necessarily increase with the number of rows. The runtime for non-interactive ptype has been shown to scale linearly with the number of unique values $U$, handling around $10K$ unique values per second \cite{ptype}. This makes the ptype AI assistant feasible in practice. It would be possible to further improve its performance by parallelizing the computations. For instance, the calculation of the probabilities assigned for unique data values can be calculated independently.

\parag{CleverCSV}
The CleverCSV AI assistant uses the non-interactive algorithm of
\cite{ccsv} to identify the optimal formatting dialect for a given CSV file,
$\Din$. The possible dialects are generated by CleverCSV prior to the
optimization, and are based on the set, $\mathcal{C}$, of unique characters in
the file. The optimization proceeds by computing for each dialect a ``pattern
score'' that captures how regular the structure of the parsed data is (i.e.,~whether
the resulting table has the same number of cells in each row), and a
``type score'' that captures the proportion of cells in the parsed file with
an identifiable data type. The product of these two scores forms the objective
function to be maximized.  Since the type score is in the range $[0, 1]$,
computing it can be skipped for dialects with a small value for the pattern
score (see \cite{ccsv}).

We distinguish three components of CleverCSV: constructing
potential dialects, computing pattern scores, and computing type scores.
Constructing the dialects can be done naively in
$O(|\mathcal{C}|)$ time, with $|\mathcal{C}|$ denoting the number of
elements in the set $\mathcal{C}$. In \cite{ccsv} two pruning steps are
discussed to remove unlikely dialects, which increases the
complexity for constructing potential dialects to $O(|\Din|\cdot
|\mathcal{C}|^3)$. Theoretically, the number of potential dialects is on the
order of $|\mathcal{C}|^3$. Computing the pattern score for each dialect is
linear in the size of the input file, $O(|\Din|)$.  If we write
$\mathcal{T}$ for the set of data types in the type score,
then computing this score can be done in $O(|\Din|\cdot
|\mathcal{T}|)$, as the number of cells in the parsing result is linear in
the size of the input.  Combining these results gives a worst-case
runtime complexity of $O(|\Din|\cdot |\mathcal{T}|\cdot
|\mathcal{C}|^3)$. However, as discussed in \cite{ccsv}, the number of
potential dialects is in practice proportional to $|\mathcal{C}|$, giving a
practical runtime complexity of $O(|\Din|\cdot
|\mathcal{T}|\cdot|\mathcal{C}|)$. The median runtime for the files in
Section~\ref{sec:eval-quantitative} is 0.018 seconds.

Human interaction with the CleverCSV AI assistant provides constraints on the
dialects considered for the file. By storing the value of the
objective function for each dialect in a lookup table, interactions
with the AI assistant need only update the allowed dialects in this table,
resulting in interactions that are linear in the number of dialects.

\parag{ColNet}
The non-interactive version of ColNet trains a CNN classifier for each (relevant) semantic type in the knowledge graph.
The training is split into two phases \cite{colnet}: pre-training and fine-tuning. The pre-training is performed using the
information from the knowledge graph (typically a large set of samples) while the fine-tuning is computed with the data
from the column to be annotated (typically a small set of samples).

As described in \cite{colnet}, the classifiers were implemented in Tensforflow and the pre-trained phase for each classifier
was completed within 2 minutes on a workstation with Xeon CPU E5-2670. The computation time for the fine-tuning phase was in the order of seconds. The interactive version of ColNet relies on the same training phases, where the pre-training can be run offline for each knowledge graph. Fine-tuning only needs to be run before the first human interaction and it is done using a sample drawn from the input data. The sample size can thus be adapted to meet given performance goals. For the cases of very large knowledge graphs, one could also focus only on a subset of relevant types.

The constraints used by ColNet, as described in Section \ref{sec:aia-colnet}, directly affect the score associated to a semantic type for the involved sample and has an impact on the overall score of a semantic type for the column. Constraints obtained during the user interaction could also be used to further fine-tune the involved classifiers and thus adapt their scores. This would affect the performance, but could potentially improve the quality of recommendations.

\section{Outlier AI assistant}
\label{sec:aia-outlier}

The AI assistants discussed so far are examples of tools based on sophisticated machine learning
methods. Such tools allow the analyst to tackle the most challenging data wrangling tasks.
However, data analysts also regularly need to complete more mundane tasks, such as identifying
outlier values based on standard deviation, removing exact duplicates or correcting simple typos.
Such mundane tasks would typically be done without dedicated tool support. However, the fact that
AI assistants are very easy to build makes it practical to develop dedicated interactive tools
to support mundane tasks that are based on a simple algorithms.

\subsubsection*{Formal definition}
We first discuss a simplified formal model of an AI assistant for removing outlier values based on
the $m$-sigma rule. Given a sequence of values $x_1, \ldots, x_n$ with a mean $\overline{x}$
and a standard deviation $\sigma$, the assistant identifies values outside of the interval
$(\overline{x}-m\sigma, \overline{x}+m\sigma)$ for a multiplier $m$ specified by the analyst.
It then offers the values outside of the range to
the analyst who can choose which of those should be removed from the dataset. For simplicity,
we describe a version of the assistant where the input is a sequence of values, corresponding
to a data table with a single column.

The outlier assistant is not optimization-based. It offers
potential outliers as a result of the $\choices_\Din$ operation. The user can then choose values
to be removed. A human interaction $H$ is thus a set of values selected by the user. The
expressions are likewise just sets of values to be removed. The $\best_\Din$ operation does not
perform any inference and simply returns the values selected by the user. The $f$ operation then
actually removes the values from the dataset. Assuming $O$ is a set of outliers $o_1, \ldots, o_n$
such that $o_i\in\Din$ and $o_i\leq \overline{x}-m\sigma$ or $o_i\geq \overline{x}-m\sigma$,
the assistant is defined as:
\begin{equation*}
\begin{array}{l}
f(e, \Din) = \{ x_i \in \Din \;|\; x_i \notin e \}\\[0.25em]
\best_\Din(H) = H\\[0.25em]
\choices_{\Din}(H) = H \cup \{ o_1 \}, \ldots, H \cup \{ o_k \}\\[0.25em]
\quad \textit{where}~o_1, \ldots, o_k = O \setminus H
\end{array}
\end{equation*}
The expression $e$ is a set of values that the user selected for removal. To apply the
expression, the $f$ function removes all values from $\Din$ that are also in $e$.
Since human interactions $H$ and expressions $e$ are the same, the $\best_\Din$ function
simply returns the human interaction $H$ it receives as an argument as the best cleaning script.
Finally, the $\choices_\Din$ operation takes previously selected values to be removed $H$.
It generates a list of choices by taking all outlier values that are not already selected,
i.e.,~$O\setminus H$, and adds each to the already selected outliers to be removed.

The value of this example is two-fold. First, it implements a simple yet practical operation
that data scientists in the real world actually use. For example, the anomaly detection in
the Tundra Traits case study discussed in \cite{nazabal2020aida} uses this approach with
an $8\sigma$ threshold. Second, the example shows the flexibility of our definition. It supports
optimization-based AI assistants, but also more manual ones such as the Outlier assistant
described here.


\subsubsection*{Removing aggregates}

To illustrate the usefulness of simple AI assistants, we developed a practical version of the
AI assistant for outlier detection based on the simple theoretical model presented above.
The assistant can be used for removing outlier rows, for example when working with datasets that
combine raw and aggregate data. This example illustrates the possibilities of
the AI assistant ecosystem. It is a simple assistant that solves a specific problem, but does so
very effectively.

The assistant takes a data table with a mix of numerical and categorical columns. It identifies rows that contain
numerical outliers (using a simple $m$-sigma rule) and collects values of categorical columns in
those rows. The user can choose any of those as conditions for filtering rows in the dataset.
The user can choose to remove all rows where a selected categorical column has a particular
value that has been found among the outlier rows.

Consider data on aviation incidents published by Eurostat\footnote{\url{https://ec.europa.eu/eurostat/web/transport/data/main-tables}}
(Table~\ref{tab:avia-data}). Each row shows the number of people injured in
accidents that involve an airplane registered in a country specified by \texttt{c\_regis} that
occurred in a country given in the \texttt{c\_geo} column. However, the dataset also contains
aggregate rows. The last row in the sample shows the total number of injuries in the EU, which
is obtained as a sum of all the other rows (some not shown). Such aggregate rows are not
uncommon in real-world datasets, and can significantly affect an analysis if they are not
identified.

To work with the data, the analyst first wants to remove the aggregate rows. When she invokes the AI
assistant for outlier detection on the aviation accidents dataset, she gets four recommendations
related to the \texttt{c\_regis} column and three recommendations related to the \texttt{c\_geo} column.
The assistant offers a choice of transformations that remove rows where \texttt{c\_regis} is \texttt{EU28}, \texttt{FR},
\texttt{CH} or \texttt{NEASA} and rows where \texttt{c\_geo} is \texttt{EU28}, \texttt{OTH} or
\texttt{FR}. With two human interactions, the analyst can choose the desired two filters and remove
all aggregates (either of the columns has a value \texttt{EU28}) from the dataset. The other choices
are not relevant, but indicate regions that are worthy of further investigation, e.g.,~France (with
higher than average number of accidents) and planes registered outside of the EU (denoted by \texttt{NEASA}).

In R or Python, the analyst could write code to identify rows with values outside of the $m$-sigma
range. She might notice the \texttt{EU28} value and write code to remove rows where \texttt{c\_geo}
or \texttt{c\_regis} are \texttt{EU28}. This is easy for a seasoned programmer, but our AI
assistant allows a non-programmer to solve the problem with two simple interactions.

In Trifacta \cite{trifacta} the analyst can use the data quality bar and histogram (automatically
displayed for each column) to locate unusual values in each column data. The ``Column Details''
window also offers a list of outlier values (identified based on proprietary chosen quantile
in each column). Based on the outlier values, the analyst can construct a filter for removing rows,
e.g.,~where the value for the \texttt{2017} column is between 15 and 25. However, Trifacta
operates on individual columns and so it is not immediately obvious that the outliers represent
aggregates with a special value in separate \texttt{c\_regis} and \texttt{c\_geo} columns.


\begin{table}
\caption{Subset of Eurostat data on aviation accidents.}
\label{tab:avia-data}
\vspace{-0.5em}
\centering
\renewcommand{\arraystretch}{1.1}
\setlength{\tabcolsep}{1em}
\begin{tabular}{llcccc}
  \toprule \texttt{c\_regis} & \texttt{c\_geo} & 2017 & 2016 & 2015 & 2014\\
  \midrule UK & CZ & 0 & 0 & 0 & 0\\
  UK & IT & 0 & 0 & 1 & 0\\
  UK & SE & 0 & 0 & 0 & 0\\
  UK & UK & 3 & 0 & 2 & 2\\
  EU28 & EU28 & 18 & 7 & 22 & 31\\
  \bottomrule
\end{tabular}
\vspace{-0.5em}
\end{table}

\section{System overview}
\label{sec:system}

AI assistants are available as an extension for the industry standard JupyterLab notebook
system. Figure~\ref{fig:ddiff-jupyter} shows the use of the datadiff
AI assistant for solving the problem discussed in Section~\ref{sec:eval-ddiff}.

As discussed in Section~\ref{sec:theory-pl}, the fact that AI assistants use a
unified interface means that a single extension provides access to a wide range of AI
assistants available in a single data analysis environment. Our support for AI assistants
utilizes the Wrattler extension \cite{wrattler} for JupyterLab. In this section, we discuss
the system architecture and implementation of the abstract interface of AI assistants.
The code for the Wrattler extension and several of the AI assistants is available at: \url{https://github.com/wrattler}.

\begin{figure}
\centering
\includegraphics[scale=0.64,trim={1cm 0.9cm 1cm 0.9cm},clip]{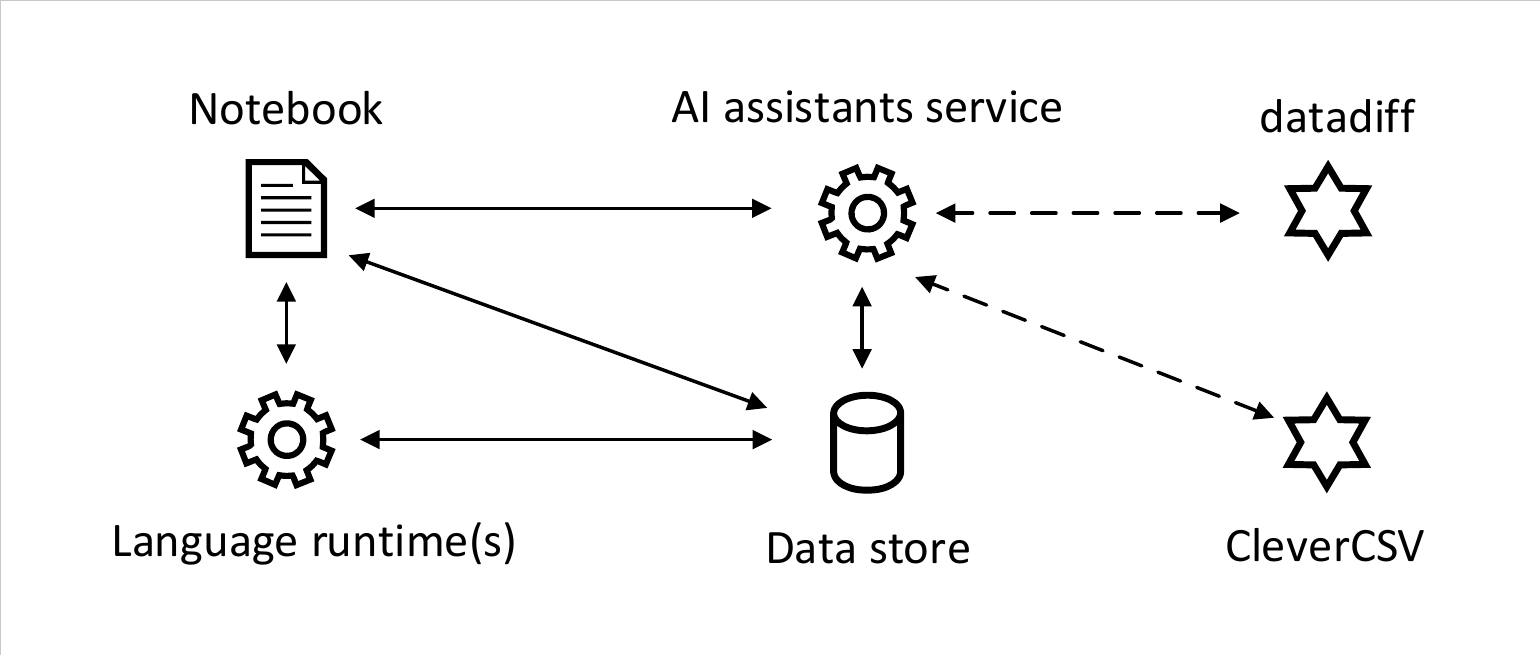}
\caption{AI assistants in Wrattler (partly adapted from \cite{wrattler}).
Wrattler keeps all data in data store on the server.
The notebook communicates with language plugins and data store via HTTP (solid lines).
We add a new service that facilitates access to AI assistants, which communicates with
individual assistants using standard input/output (dashed lines).}
\label{fig:architecture}
\end{figure}

\parag{System architecture}
Our implementation leverages Wrattler \cite{wrattler}, which extends JupyterLab with a new
kind of polyglot notebook that can contain multiple kinds of cells. The Wrattler architecture,
including the support for AI assistants, is illustrated in Figure~\ref{fig:architecture}.
Wrattler separates the notebook (running in a
web browser), from language runtimes and a data store (running on a server). Our extension implements
a language plugin for Wrattler that defines a new ``AI assistant'' cell type and facilitates
access to individual AI assistants. The new cell type uses a graphical user interface that
allows users to choose the assistant they want to invoke, as well as select the input data.
When the cell is evaluated, it invokes the AI assistant, previews the results and allows the
user to select one of the options generated by the $\choices_\Din$ operation of the
AI assistant.

\parag{Common interface and integration}
Our implementation aims to make it easy to create new AI assistants. For this
reason, AI assistants use a shared and easy-to-implement communication interface---standard
input and output---together with a simple protocol that implements the abstract definition.

Definition~\ref{def:aia} defines an AI assistant formally in terms of operations $f$,
$\best_\Din$, and $\choices_\Din$. In our implementation, AI assistants are command-line
applications that read commands (corresponding to the operations) from standard input and
respond via standard output. For example, our JupyterLab integration calls datadiff to
get completions after the ``Don't transform LLU'' constraint is selected by the analyst as follows
(\verb+>+ marks standard input, \verb+<+ marks standard output):

\begin{verbatim}[numbers=left,xleftmargin=6mm]
> reference=/temp/bb15nice.csv,input=/temp/bb14.csv
> choices
> notransform(LLU)

< Don't transform 'Urban.rural'
< notransform(LLU)/notransform(Urban.rural)
< Don't match 'Nation' and 'Urban.rural'
< notransform(LLU)/nomatch(Nation,Urban.rural)
<
\end{verbatim}

\noindent
The first three lines invoke the $\choices_\Din$ operation by specifying input data (line 1),
operation name (line 2) and past human interactions $H$ (line 3). The response generates
possible human interactions followed by a blank line. The actual implementation returns
multiple choices, but we list only the first two in the above example. Each choice consists
of a name, followed by a new human interaction. Here, the human interactions are encoded as
constraints, separated by a slash. The analyst previously selected the $\ident{notransform}(\ident{LLU})$
constraint, so the two offered human interactions include this and add one other constraint.
The first one, named ``Don't transform Urban.rural'' (line 5), is represented as two constraints
(line 6), the existing $\ident{notransform}(\ident{LLU})$ constraint and a newly added
$\ident{notransform}(\ident{Urban.rural})$ constraint. The second human interaction (lines 7-8)
similarly represents two constraints, the existing $\ident{notransform}(\ident{LLU})$ constraint
and a newly added $\ident{nomatch}(\ident{Nation},\ident{Urban.rural})$ constraint.

We chose standard input/output as our interface, because it makes it possible to implement
AI assistants in any programming language. For example, the assistants presented in this paper
have been implemented in R (datadiff), Python (CleverCSV, ptype), and F\# (Outlier).

\begin{table*}[!t]
\caption{An overview of datasets used throughout the paper and their sources.}
\label{tab:datasets}
\centering
\renewcommand{\arraystretch}{1.1}
\setlength{\tabcolsep}{0.6em}
\begin{tabular}{lllll}
  \toprule
  Name & Description & Use & Source & Size\\
  \midrule
  Broadband (2014) & UK home broadband performance & datadiff motivation & Ofcom \cite{ofcom} & 32 cols, 1971 rows\\
  Broadband (2015) & UK home broadband performance & datadiff motivation & Ofcom \cite{ofcom} & 67 cols, 2802 rows\\
  IMDB movies & Classification and rating of 100 movies & CleverCSV evaluation & Kaggle ($*$) & 44 cols, 100 rows\\
  Colors & File names and RGB color codes & CleverCSV scenario & GitHub ($\dagger$) & 11 cols, 300 rows\\
  Cylinder Bands & Cylinder bands in rotogravure printing & ptype scenario & UCI \cite{uci} & 40 cols, 512 rows\\
  Corrupted UCI (1) & Corrupted (abalone, adult, bank, car, iris) & datadiff evaluation & UCI \cite{uci} & max 15 cols, 32,561 rows\\
  Corrupted UCI (2) & Corrupted (abalone, adult, bank, car) & datadiff evaluation & UCI \cite{uci} & max 15 cols, 32,561 rows\\
  CleverCSV failures & Subset of data from Gov.uk and GitHub & CleverCSV evaluation & CleverCSV \cite{ccsv} & 255 files\\
  ptype failures & Subset of data from Gov.uk and UCI & ptype evaluation & ptype \cite{ptype} & 43 columns\\
  Aviation accidents & EU aviation accidents per year & outlier scenario & Eurostat ($\ddagger$) & 32 cols, 3469 rows\\
  \bottomrule
\end{tabular}
\begin{flushleft}
\vspace{0.5em}
\hspace{2.2em} ($*$) \url{https://github.com/alan-turing-institute/CleverCSV/blob/master/example/imdb.csv} \\
\hspace{2.2em} ($\dagger$) \url{https://github.com/victordiaz/color-art-bits-} \\
\hspace{2.2em} ($\ddagger$) \url{https://ec.europa.eu/eurostat/web/transport/data/main-tables}
\end{flushleft}
\vspace{-1em}
\end{table*}

\begin{table*}[!t]
\caption{A glossary of symbols and special identifiers used throughout the paper.}
\label{tab:symbols}
\centering
\renewcommand{\arraystretch}{1.1}
\setlength{\tabcolsep}{0.6em}
\begin{tabular}{lll}
  \toprule
  Symbol & Scope & Explanation\\
  \midrule
  $e, e^{*}$ & AI assistants & Expressions (cleaning scripts) recommended by AI assistants; $e^{*}$ denotes the best script \\
  $\Din, \Dout$ & AI assistants & Input dataset $\Din$ and output dataset $\Dout$ \\
  $H, H_0$ & AI assistants & Past human interactions with the AI assistant; $H_0$ denotes no prior interaction\\
  $f(e,\Din)$ & AI assistants & Operation that applies the expression $e$ (cleaning script) to the input dataset $\Din$ \\
  $\best_{\Din}(H)$ & AI assistants & Operation that recommends the best expression for a given input, respecting past interactions \\
  $\choices_{\Din}(H)$ & AI assistants & Operation that generates a sequence of options the analyst can choose from \\
  \midrule
  $Q_H(\Din, e), Q$ & Optimization & Objective function that assigns a score to an expression, w.r.t. past interaction ($Q_H$)\\
  $E_H, E$ & Optimization & Set of permitted expressions; $E_H$ is restricted with respect to past human interaction \\
  $p_{H}(\Din\,|\,e)$ & Probabilistic & Likelihood of the input data $\Din$ given an expression $e$, w.r.t. past human interaction\\
  $p_{H}(e)$ & Probabilistic & Distribution representing prior beliefs about probabilities of expressions\\
  \midrule
  $P$ & datadiff & A single patch that can be applied to a column of the dataset\\
  $c$ & datadiff, CleverCSV & A single constraint that can be added to $H$ in order to influence the inference\\
  $\ident{valid}_H$ & datadiff, CleverCSV & A predicate that determines if a patch or type respects past interactions (constraints)\\
  $\tau$ & ptype & Inferred primitive type such as Boolean, integer, floating-point number, date or string\\
  $\sigma$ & ColNet & Inferred semantic type from a knowledge graph such as \ident{dbo:Company} \\
  $S$ & ColNet & Set of sample values, drawn from a column of the input dataset\\
  $p_{S}^{\sigma}$  & ColNet & Score of a sample $S$ for a given semantic type $\sigma$ in non-interactive mode\\
  $q_{S, H}^{\sigma}$  & ColNet & Score of a sample $S$ for a given semantic type $\sigma$; w.r.t past interactions\\
  \bottomrule
\end{tabular}
\end{table*}

\newpage
~

\end{document}